\documentclass[aps,amsmath,amssymb,superscriptaddress]{revtex4}
\usepackage{graphicx}

\begin{document}
\title{Snapshots of the retarded interaction of charge carriers with ultrafast fluctuations in cuprates}

\author{S. Dal Conte}
\affiliation{IFN-CNR, Dipartimento di Fisica, Politecnico di Milano, 20133 Milano, Italy}

\author{L. Vidmar}
\affiliation{Department of Physics and Arnold Sommerfeld Center for Theoretical Physics, Ludwig-Maximilians-Universit\"at M\"unchen, D-80333 M\"unchen, Germany}
\affiliation{J. Stefan Institute, 1000 Ljubljana, Slovenia}

\author{D. Gole\v{z}}
\affiliation{J. Stefan Institute, 1000 Ljubljana, Slovenia}

\author{M. Mierzejewski}
\affiliation{Institute of Physics, University of Silesia, 40-007 Katowice, Poland}

\author{G. Soavi}
\affiliation{IFN-CNR, Dipartimento di Fisica, Politecnico di Milano, 20133 Milano, Italy}

\author{S. Peli}
\affiliation{i-LAMP (Interdisciplinary Laboratories for Advanced Materials Physics), Universit\`a Cattolica del Sacro Cuore, Brescia I-25121, Italy}
\affiliation{i-LAMP (Department of Physics, Universit\`a degli Studi di Milano, Italy}

\author{F. Banfi}
\affiliation{i-LAMP (Interdisciplinary Laboratories for Advanced Materials Physics), Universit\`a Cattolica del Sacro Cuore, Brescia I-25121, Italy}
\affiliation{Department of Mathematics and Physics, Universit\`a Cattolica del Sacro Cuore, Brescia I-25121, Italy}

\author{G. Ferrini}
\affiliation{i-LAMP (Interdisciplinary Laboratories for Advanced Materials Physics), Universit\`a Cattolica del Sacro Cuore, Brescia I-25121, Italy}
\affiliation{Department of Mathematics and Physics, Universit\`a Cattolica del Sacro Cuore, Brescia I-25121, Italy}

\author{R. Comin}
\affiliation{Department of Physics and Astronomy, University of British Columbia, Vancouver, BC V6T 1Z1, Canada}
\affiliation{Quantum Matter Institute, University of British Columbia, Vancouver, BC V6T 1Z4, Canada}

\author{B.M. Ludbrook}
\affiliation{Department of Physics and Astronomy, University of British Columbia, Vancouver, BC V6T 1Z1, Canada}
\affiliation{Quantum Matter Institute, University of British Columbia, Vancouver, BC V6T 1Z4, Canada}

\author{L. Chauviere}
\affiliation{Department of Physics and Astronomy, University of British Columbia, Vancouver, BC V6T 1Z1, Canada}
\affiliation{Quantum Matter Institute, University of British Columbia, Vancouver, BC V6T 1Z4, Canada}
\affiliation{Max Planck Institute for Solid State Research, Heisenbergstrasse 1, D-70569 Stuttgart, Germany}

\author{N.D. Zhigadlo}
\affiliation{Laboratory for Solid State Physics, ETH, 8093 Z\"urich, Switzerland}

\author{H. Eisaki}
\affiliation{Nanoelectronics Research Institute, National Institute of Advanced Industrial Science and Technology, Tsukuba, Ibaraki 305-8568, Japan}

\author{M. Greven}
\affiliation{School of Physics and Astronomy, University of Minnesota, Minneapolis, Minnesota 55455, USA}

\author{S. Lupi}
\affiliation{CNR-IOM Dipartimento di Fisica, Universit\`a di Roma ÓLa SapienzaÓ P.le Aldo Moro 2, 00185 Rome, Italy}

\author{A. Damascelli}
\affiliation{Department of Physics and Astronomy, University of British Columbia, Vancouver, BC V6T 1Z1, Canada}
\affiliation{Quantum Matter Institute, University of British Columbia, Vancouver, BC V6T 1Z4, Canada}

\author{D. Brida}
\affiliation{IFN-CNR, Dipartimento di Fisica, Politecnico di Milano, 20133 Milano, Italy}
\affiliation{Department of Physics and Center for Applied Photonics,
University of Konstanz, 78457 Konstanz, Germany}

\author{M. Capone}
\affiliation{CNR-IOM Democritos National Simulation Center}
\affiliation{Scuola Internazionale Superiore di Studi Avanzati (SISSA), Via Bonomea 265, 34136 Trieste, Italy}

\author{J. Bon\v{c}a}
\affiliation{J. Stefan Institute, 1000 Ljubljana, Slovenia}
\affiliation{Faculty of Mathematics and Physics, University of Ljubljana, 1000 Ljubljana, Slovenia}

\author{G. Cerullo}
\affiliation{IFN-CNR, Dipartimento di Fisica, Politecnico di Milano, 20133 Milano, Italy}

\author{C. Giannetti}
\affiliation{i-LAMP (Interdisciplinary Laboratories for Advanced Materials Physics), Universit\`a Cattolica del Sacro Cuore, Brescia I-25121, Italy}
\affiliation{Department of Mathematics and Physics, Universit\`a Cattolica del Sacro Cuore, Brescia I-25121, Italy}

\begin{abstract}
\textbf{One of the pivotal questions in the physics of high-temperature superconductors is whether the low-energy dynamics of the charge carriers is mediated by bosons with a characteristic timescale. This issue has remained elusive since electronic correlations are expected to dramatically speed up the electron-boson scattering processes, confining them to the very femtosecond timescale that is hard to access even with state-of-the-art ultrafast techniques.
Here we simultaneously push the time resolution and the frequency range of transient reflectivity measurements up to an unprecedented level that enables us to directly observe the $\sim$16 fs build-up of the effective electron-boson interaction in hole-doped copper oxides. This extremely fast timescale is in agreement with numerical calculations based on  the $t$-$J$ model and the repulsive Hubbard model, in which the relaxation of the photo-excited charges is achieved via inelastic scattering with short-range antiferromagnetic excitations.}
\end{abstract}
\maketitle

After almost 30 years of intensive experimental and theoretical efforts to understand the origin of high-temperature superconductivity in copper oxides, a consensus about the microscopic process responsible for the superconducting pairing is still lacking.
The large Coulomb repulsion $U\gg$1~eV between two electrons occupying the same lattice site is believed to have fundamental consequences for the normal state of these systems \cite{Lee2006}, and it is not clear whether a BCS-like bosonic glue that mediates the electron interactions and eventually leads to pairing can still be defined \cite{Anderson2007,Scalapino2012,Monthoux2007}. 
The fundamental issue can be reduced to the question whether the electronic interactions are essentially instantaneous, or whether the low-energy physics, including superconductivity, can be effectively described in terms of  interactions among the fermionic charge carriers mediated by the exchange of bosons. 
The problem can be rationalized by considering the Hubbard model, in which the instantaneous virtual hopping of holes into already occupied sites (with an energy cost of $U$) inherently favors an antiferromagnetic (AF) coupling 
$J$=$4 t_h^2 / U$ between neighbouring sites, where $t_h$ is the nearest-neighbour hopping energy. As a consequence, antiferromagnetic fluctuations with a high-energy cutoff of $2J$$\ll$$U$ naturally emerge as a candidate for mediating the low-energy electronic interactions, on a characteristic retarded timescale on the order of $\hbar$/2$J$.

In principle, time-resolved optical spectroscopy \cite{Orenstein2012} may be used to prove the existence of an effective retarded boson-mediated interaction, provided that: i) the temporal resolution is on the order of the inverse bosonic fluctuation scale (e.g. $\hbar$/2$J$ for AF fluctuations); ii) the optical properties are probed over a sufficiently broad frequency range, to obtain direct information on the dynamics of the electron-boson coupling. 
Here we develop a transient reflectivity experiment \cite{Brida2010}, based on extremely short broadband pulses (9-13 fs) in the infrared-visible spectral region (0.75-2.4 eV) (see Methods), that is most sensitive to the variations of the electron-boson scattering rate. The goal is to investigate possible retardation effects of the electron-boson interaction that were inaccessible in previous measurements with $\sim$100 fs time resolution \cite{dalconte12}. The qualitative idea of the experiment is to use an ultrashort light pulse to impulsively increase the kinetic energy of a small fraction of the charge carriers (photo-excited holes) in doped cuprates. A second broadband pulse probes the instantaneous optical scattering rate by inducing boson-assisted optical transitions of the charge carriers in the conduction band. As long as the number of photo-excited holes is small as compared to the intrinsic doping of the system (see Methods), we can assume that we are probing the average scattering rate experienced by the holes which have not been directly excited by the pump pulse. Since the optical scattering rate is proportional to the energy stored in the bosonic fluctuations, we can monitor the time needed to complete the energy exchange from the photo-excited carriers to the bosonic bath.

The dynamics of the optical scattering rate after the impulsive excitation can be directly monitored by probing the damping of the infrared plasma edge of a prototypical superconducting cuprate. This is easily shown by inspection of the room-temperature equilibrium reflectivity, $R_{eq}(\omega)$, of a Bi$_2$Sr$_2$Y$_{0.08}$Ca$_{0.92}$Cu$_2$O$_{8+\delta}$ (YBi2212) crystal \cite{Eisaki2004}. The doping level (hole concentration $p$$\simeq$0.13, $T_c$$\simeq$83 K) is close to optimal ($p$$\simeq$0.16, $T_c$$\simeq$96 K), and we estimate the opening of the pseudogap to occur at $T^*$$\simeq$240 K \cite{Cilento2014}. As shown in Figure 1a, the infrared region is dominated by a broad plasma edge,  which can be effectively reproduced by an Extended Drude Model (EDM) \cite{Heumen2009} described by $4\pi\sigma_D(\omega)$=$\omega^2_p/[\gamma(\omega,T)$-$i\omega(1$+$\tilde{\lambda}(\omega,T))]$ where $\sigma_D(\omega)$ is the EDM optical conductivity, $\omega_p$ the plasma frequency and 1+$\tilde{\lambda}$ the renormalization factor of the effective mass of the carriers. The optical scattering rate, $\gamma(\omega,T)$, controls the damping of the plasma edge and it is, in principle, a function of the frequency and temperature, as a consequence of the coupling to bosonic fluctuations whose occupation number is regulated by the Bose-Einstein statistics \cite{Heumen2009}. Nevertheless, in the energy window probed in the experiment, $\gamma(\omega,T)$ varies less than 10$\%$ as a function of frequency (see Supplementary Information), therefore we can safely discuss the time-resolved results considering its asymptotic ($\omega$$\rightarrow$$\infty$) value, i.e., $\gamma_{\infty}(T)$. If the total scattering rate is enhanced ($\delta\gamma_{\infty}$$>$0) by any effect, this leads to a further damping of the edge and to the evolution of $R_{eq}(\omega)$ shown in Figure 1a.  An isosbestic point is found at the frequency $\tilde{\omega}$$\simeq$1.1 eV, across which the reflectivity variation changes from negative to positive. 

\begin{figure}
\includegraphics[keepaspectratio,clip,width=0.9\textwidth]{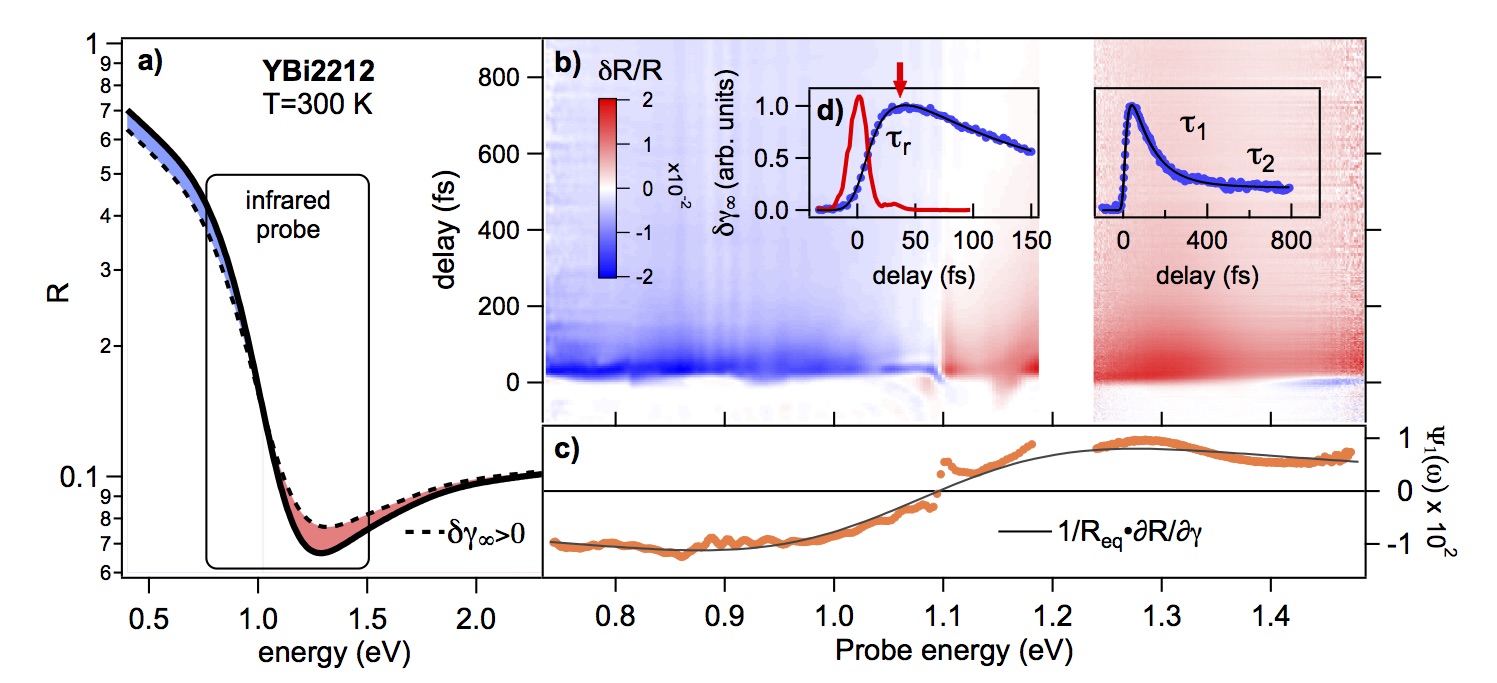}
\caption{Equilibrium and non-equilibrium optical properties of YBi2212. \textbf{a)} The normal state reflectivity (black line) of the YBi2212 crystal is reported. The dashed line represents the reflectivity calculated by increasing the electron-boson scattering rate, $\gamma_{\infty}$, in the extended Drude model. \textbf{b)} The ultrafast dynamics of $\delta R/R$($\omega$,t), measured across the isosbestic point $\tilde{\omega}$, is reported. The measurements have been performed with an incident pump fluence of 0.7 mJ/cm$^2$. The linearity of the optical response has been checked in the 0.2-2 mJ/cm$^2$ fluence range.  \textbf{c)} The first eigenfunction, $\Psi_1(\omega)$, obtained through the SVD (see Supplementary Information) is displayed (yellow dots). SVD consists of decomposing $\delta R$($\omega$,t)/$R_{eq}(\omega)$$\simeq$$\sum_k$$\Psi_k(\omega)\delta \phi_k$(t). The fit obtained by increasing the total scattering rate, i.e., $\Psi_1(\omega)$=1/$R_{eq}(\omega)$$\cdot$$\partial R$/$\partial \gamma$$(\omega)$, is reported as a black line. \textbf{d)} The time-dependent scattering rate variation, $\delta \gamma_{\infty}$(t)=$\delta \phi_1$(t), is reported (blue dots). The black line is the fit to the data of the function (1-exp(-t/$\tau_r$))$\cdot$(I$_1$exp(-t/$\tau_1$)+I$_2$exp(-t/$\tau_2$)) convoluted with a gaussian pulse accounting for the experimental time-resolution. The red line is the measured cross-correlation trace between the pump and probe pulses, that sets the time-resolution to 19$\pm$2 fs. The red arrow highlights the time delay ($\sim$40 fs) at which the maximal $\delta \gamma_{\infty}$ variation is measured.} 
\label{fig1}
\end{figure}
In Figure 1b we report the dynamics of the relative reflectivity variation, $\delta R$($\omega$,t)/$R_{eq}(\omega)$, measured at $T$=300 K on YBi2212 as a function of the delay t between the pump and probe pulses. The sub-10 fs pump pulse is set to 2.1 eV central energy, in order to avoid the spurious signals measured when pumping below the Drude absorption edge \cite{Novelli2013}. The probe spans the energy range across $\tilde{\omega}$, that is more sensitive to the possible variations of  $\gamma_{\infty}(T)$, as already shown in Fig. 1a. One key result of this experiment is that, for each given time delay t from the very first femtoseconds up to the picosecond timescale,  $\delta R$($\omega$,t)/$R_{eq}(\omega)$ can be accurately reproduced exclusively by increasing $\gamma_{\infty}$. 
This is corroborated by the quantitative analysis through Singular Value Decomposition (SVD) (see Supplementary Information), confirming that 97\% of the signal is reproduced by factorizing the reflectivity variation in $\delta R$($\omega$,t)/$R_{eq}(\omega)$=1/$R_{eq}(\omega)$$\cdot$$\partial R$/$\partial \gamma$$(\omega)$$\delta\gamma_{\infty}$(t) where $\delta\gamma_{\infty}$(t) contains the dynamics of the scattering rate variation.

Remarkably, the maximum variation of $\delta \gamma_{\infty}$(t) is measured at a finite delay time ($\sim$40 fs) from the excitation, as shown in Figure 1d. The intrinsic build-up time of the $\delta \gamma_{\infty}$(t) signal ($\tau_r$=16$\pm$3 fs) is extracted by fitting the dynamics with the simple function (1-exp(-t/$\tau_r$))$\cdot$(I$_1$exp(-t/$\tau_1$)+I$_2$exp(-t/$\tau_2$)) convoluted with a gaussian pulse accounting for the experimental resolution (pump-probe cross-correlation with a full width at half maximum of 19$\pm$2 fs), that has been carefully checked through the Cross Frequency-Resolved Optical Gating method (see Supplementary Information). This result demonstrates that the energy exchange between the photo-excited holes and the boson bath takes a non-zero time of the order of 15-20 femtoseconds. The two further decay scales ($\tau_1$$\sim$90 fs and $\tau_2$$\sim$1 ps), measured in the $\delta R$($\omega$,t) signal, are related to the subsequent coupling with the optical buckling and breathing phonons ($\tau_1$) and with the rest of the lattice vibrations ($\tau_2$) \cite{perfetti07,gadermaier10,dalconte12}. 

\begin{figure}
\includegraphics[width=0.8\textwidth]{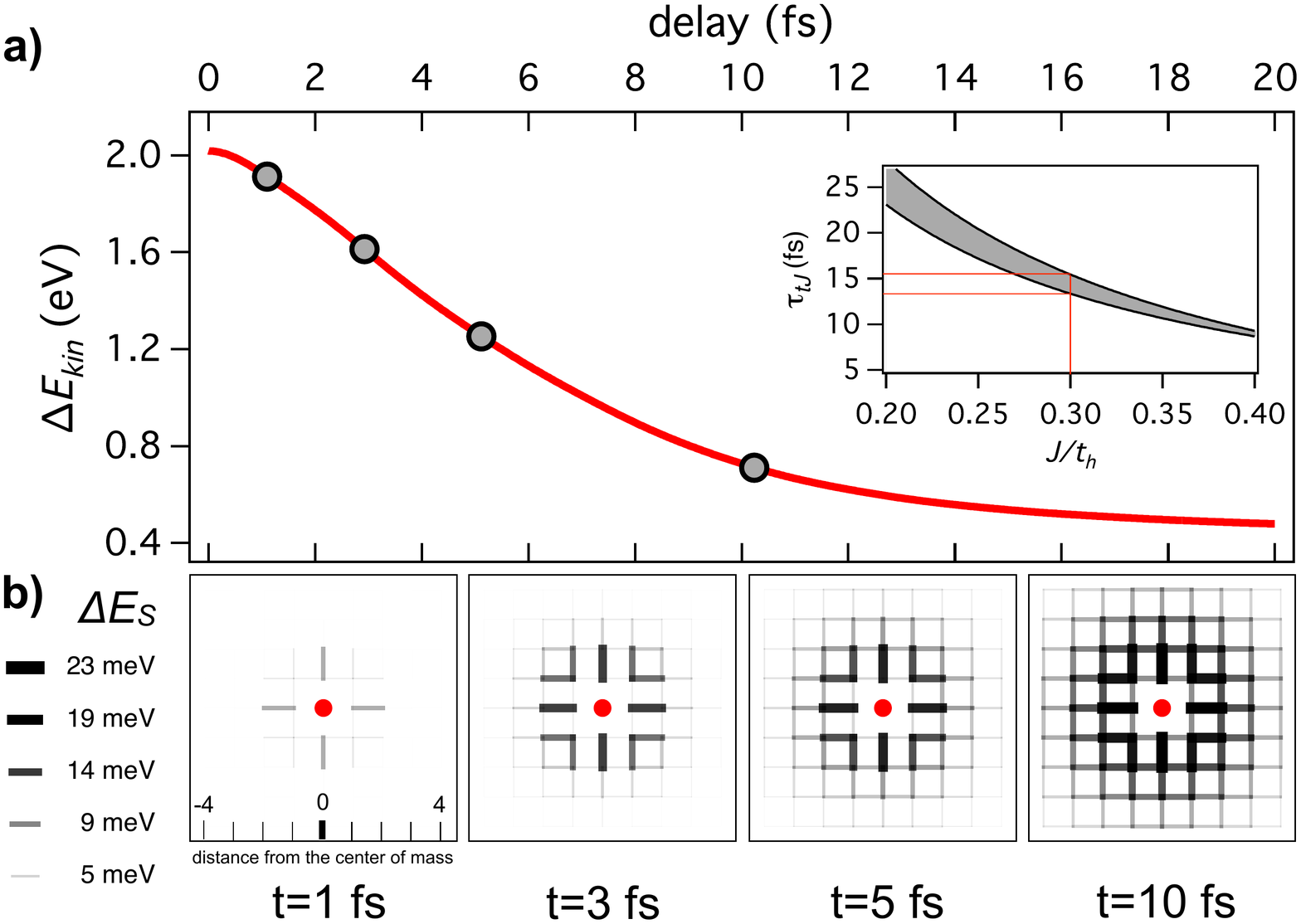}
\caption{
Microscopic calculation of the energy transfer from a photo-excited hole to local antiferromagnetic bonds through the out-of-equilibrium $t$-$J$ model (see Methods for further details). 
 \textbf{a)}
The time evolution of the hole kinetic energy variation, $\Delta E_{\rm kin}$(t).
In the main panel we set $J/t_h$=$0.3$, where $t_h$=$360$~{\rm meV} represents the hopping amplitude. 
The inset displays the range of the characteristic relaxation time $\tau_{tJ}$ at different values of $J$/$t_h$, after a finite-size scaling (see Supplementary Information). 
\textbf{b)}
Snapshots of the relative energy increase of the local antiferromagnetic bonds $\Delta E_{\rm s}($t$,{\bf r})$ on a 9$\times$9 square lattice around the photo-excited hole (red circle).
The thickness of the black segments is proportional to the energy stored in each bond.
The grey dots in \textbf{a)} indicate delay times of the snapshots.}
\label{fig_tj}
\end{figure}

The present high time-resolution transient reflectivity experiment demonstrates that indeed the electron dynamics in the doped cuprates can be captured by an effective model in which the charge carriers exchange energy with boson excitations on an very fast (but non-zero) timescale. Considering that the intrinsic  time ($\hbar/t_h$$\sim$2 fs) associated with the Cu-O-Cu hole hopping process inevitably leads to the creation of local spin excitations, AF fluctuations are a candidate as the dominant mediators of the hole interactions on this ultrafast timescale.
As a further support of our assignment, neutron and x-ray scattering experiments \cite{Dahm2009,letacon2011,Fujita2012,dean2013,letacon2013,dean2013b} have revealed 
a rich magnetic dynamics in hole-doped cuprates, characterized by resonance modes at 50-60 meV and a very broad spectrum with a cutoff of the order of the bandwidth $2J$$\simeq$240 meV found for the AF parent compounds. Similar bosonic-fluctuation spectra have been extracted from equilibrium optical spectroscopy \cite{Heumen2009}.
On the other hand, we can rule out a possible photo-induced coherent rearrangement of the oxygen atoms, that has been demonstrated on the 20 fs timescale \cite{Singla2013}, since: i) no significant variation of the charge transfer excitations ($\sim$2 eV), that are the most sensitive to a rearrangement of the oxygen orbitals, is observed (see Supplementary Information); ii) no coherent oscillation is detected in the $\delta R$($\omega$,t)/$R_{eq}(\omega)$ signal at any wavelength. Even though a single incoherent scattering process with optical phonons could be extremely fast (1/4 of the period, i.e., $\sim$20 fs for the highest energy modes at 70 meV), several scattering processes are necessary to complete the energy transfer to the phonons, as will be shown later in the case of the conventional superconductor MgB$_2$.

\begin{figure}
\includegraphics[width=1\textwidth]{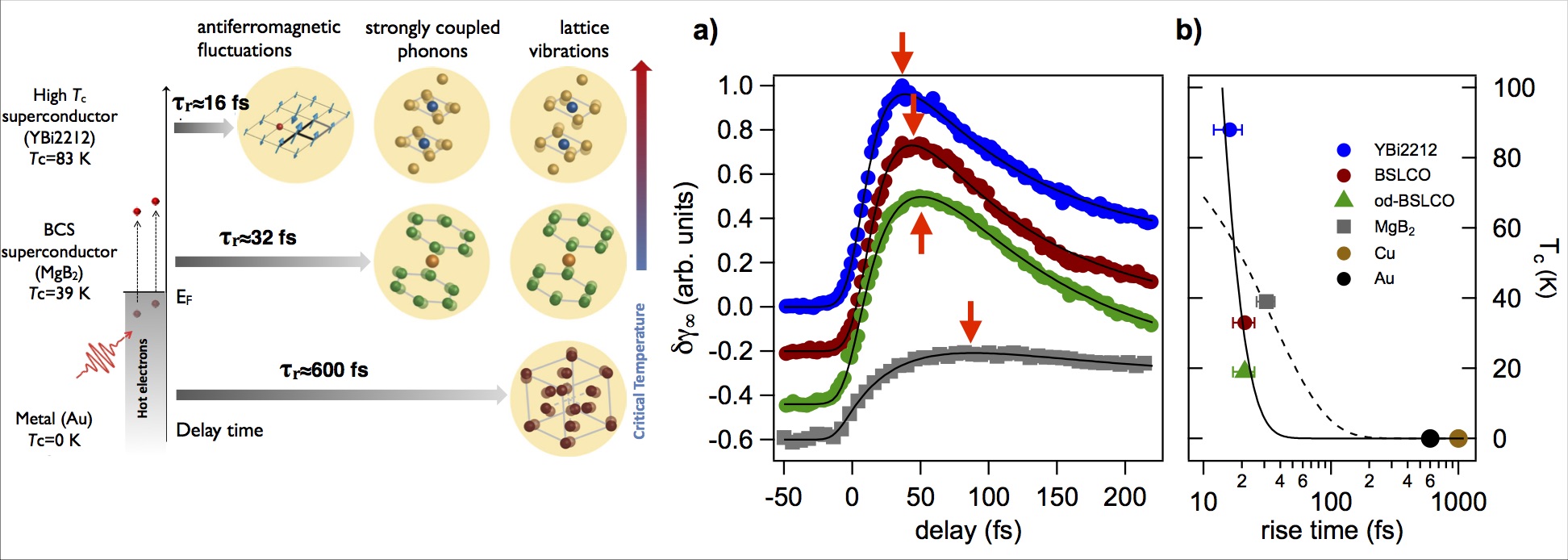}
\caption{
Ultrafast dynamics of the electron-boson interaction. The different relaxation dynamics in high-$T_c$
 superconductor and in conventional superconductors are sketched in the left panel. \textbf{a)} The dynamics of $\delta \gamma_{\infty}$(t) is shown for different superconducting systems (YBi2212, blue circles; BSLCO, red circles; od-BSLCO, green circles; MgB$_2$, grey squares). The red arrows highlight the time delay corresponding to the maximal value of $\delta \gamma_{\infty}$(t) after the excitation. \textbf{b)} The critical temperatures of the systems considered is reported as a function of the build-up time ($\tau_r$) of $\delta \gamma_{\infty}$(t), measured through the time-resolved optical spectroscopy. The black lines are the critical temperatures calculated by the Eliashberg-like expression $T_c$=0.83$\tilde{\Omega}$exp[-1.04(1+$\lambda_b$)]/[$\lambda_b$-$\mu^*(1+0.62\lambda_b)$] \cite{Allen1975}, fixing $\mu^*$=0 and assuming a generic relation between the electron-boson energy exchange rate, 1/$\tau_r$, in cuprates and the total coupling constant $\lambda_b$ and $\tilde{\Omega}$=200 meV (solid line). The dashed line represents the critical temperatures calculated for $\tilde{\Omega}$=70 meV and $\lambda_b$=1.1 compatible with the phonon-mediated pairing in MgB$_2$ \cite{Kong2001}.}
\label{fig4}
\end{figure}
The role of short-range AF correlations as the fastest effective bosons mediating the charge dynamics is corroborated by studying the non-equilibrium dynamics of the photo-excited holes in the $t$-$J$ Hamiltonian \cite{mierzejewski2011,Golez2013} (see Methods), which is the minimal model that fully retains the dynamics at the energy scale $J$.
Since the rate of the photo-excitation of holes in the system is very low (see Methods), we conjecture that the key relaxation mechanism at very short times corresponds to the creation of high-energy AF excitations in the close proximity of the photo-excited holes. We model the photo-excitation of a hole immersed in the AF background by instantaneously raising its kinetic energy, i.e., $\Delta E_{\rm kin}$(t)=2 eV. The relaxation dynamics is calculated by numerically integrating the time-dependent Schr\"odinger equation, without any assumption of quasi-thermal intermediate states, as usually described by effective electronic and bosonic temperatures. The dynamics of $\Delta E_{\rm kin}$(t) after the quench is reported in Fig. 2a. While the total energy remains constant during the time evolution, the observed relaxation of $\Delta E_{\rm kin}$(t) is compensated by the simultaneous increase of the energy stored in the AF background. Figure 2b displays snapshots of the excess spin energy density, $\Delta E_{\rm s}($t$,{\bf r})$, in the vicinity of the photo-excited hole. We note that the excitation process involves only the neighbouring bonds, demonstrating that short-range AF correlations are sufficient for describing the relaxation dynamics.
The characteristic relaxation time, $\tau_{tJ}$, of the energy transfer to AF fluctuations is governed by the ratio $J$/$t_h$, as shown in the inset of Fig. 2a.
Setting $J$/$t_h$=$0.3$ to a realistic value and performing a finite-size scaling analysis (see Supplementary Information), we obtain $\tau_{tJ}$$\sim$15~fs, that is in notable quantitative agreement with the experimental observations.
We remark that, even though the $t$-$J$ model cannot account for the optical properties at the energy scale $U$, the description of the low-energy electron dynamics through an effective charge-boson coupling is also supported by the generic single-band Hubbard model at equilibrium. In the Supplementary Material we use Dynamical Mean Field Theory (DMFT) to calculate the temperature-dependent optical conductivity of a hole-doped system ($p$=0.16). We demonstrate that the effect of an energy increase (in the form of a $\delta$$T$ variation) can be rationalized as an increase of the number of bosonic fluctuations at the energy scale $J$ without any significant change of spectral weight at the scale $U$.

Finally, valuable insight into the general relation between the electron-boson coupling and the dynamics of $\delta \gamma_{\infty}$(t) is provided by the results obtained on different families of cuprates and on more conventional superconductors (see Fig. 3). We repeated the same experiment on optimally doped (BSLCO, $x$=0.4, $T_c$=33 K) and overdoped (od-BSLCO, $x$=0.2, $T_c$=19 K) single Cu-O layer Bi$_2$Sr$_{2-x}$La$_x$CuO$_6$ crystals. The time necessary for attaining the maximal variation of $\gamma_{\infty}$ is non-zero ($\tau_r$=21$\pm$4 fs for BSLCO and $\tau_r$=22$\pm$4 fs for od-BSLCO), slightly longer than for YBi2212. These results demonstrate that the coupling to bosonic fluctuations on the ultrafast timescale is a general property of the normal state of hole-doped cuprates even at large hole-doping concentrations.
Qualitatively different results are obtained on MgB$_2$, that is considered the conventional phonon-mediated superconducting system with the highest critical temperature (T$_c$=39 K). Here, the maximum variation of the scattering rate after the impulsive excitation is significantly delayed ($\sim$90 fs, as shown in Fig. 3a) and $\tau_r$=32$\pm$3 fs is measured. This result demonstrates that, in the systems in which the superconducting pairing is mediated by high-energy ($\omega_0$$\sim$70 meV) optical phonons \cite{Kong2001}, the electron-phonon energy exchange process is completed on a timescale that is few times the inverse energy scale of the phonons involved. Furthermore, a single slower exponential decay ($\tau_1$=500 fs) is measured in the $\delta R$($\omega$,t) signal of MgB$_2$, corresponding to the energy exchange with the rest of the lattice. On conventional non-superconducting metals, like Cu and Au, time-resolved measurements have been widely applied \cite{DellaValle2012,ElsayedAli1987}, evidencing an electron-phonon coupling on the order of $\sim$600 fs in Au and $\sim$1 ps in Cu. Fig. 3 summarizes all the results obtained in both conventional and unconventional superconductors, demonstrating a universal relation between the energy scale of the boson modes mediating the electron interactions in the normal state and the timescale of the dynamics of the electron-boson scattering rate after the pump excitation. Intriguingly, the dependence of the critical temperature of the system on the experimental $\tau_r$ follows the general and qualitative trend of the approximate solution of the McMillan's equation \cite{Allen1975}, once we assume a direct relation \cite{Allen1987} between the measured 1/$\tau_r$ and the total coupling constant $\lambda_b$=2$\int$$\Pi(\Omega)$/$\Omega$d$\Omega$, where $\Pi(\Omega)$ is the coupling function to antiferromagnetic fluctuations ($I^2$$\chi(\Omega)$) for copper oxides and to phonons ($\alpha^2$$F(\Omega)$) for MgB$_2$ and metals.

Although we do not exclude that the unmediated charge-charge interactions could influence the relaxation of very high-energy excitations (i.e. those involving the doublon formation as a consequence of a charge-transfer transition), our results are consistent with a picture in which, close to the optimal hole concentration, the dynamics of the charge carriers in the conduction band can be effectively described in terms of the interaction with short-range antiferromagnetic fluctuations on a timescale ($\sim$16 fs) on the order of few times $\hbar$/2$J$. This extremely fast dynamics has been confirmed by very recent non-equilibrium DMFT calculations \cite{Eckstein2014}. We stress that, even in this picture, $U$ is the crucial interaction since it determines the dynamics through the  strength of the spin-charge coupling, $J$=$4t_h^2$/$U$. 

Our results represent a new time-domain benchmark for realistic microscopic models of the pairing in high-temperature superconductors and pave the way to unveiling the pairing mechanisms in a wealth of intriguing systems, like different families of cuprates, iron-based superconductors \cite{Hinks2009,Mazin2010} and all the materials that are on the verge of a (ferro-)antiferro-magnetic phase transition \cite{Monthoux2007}. Furthermore, we provide a platform for the next generation of time-resolved experiments aimed at addressing the role of charge-fluctuations, the dual itinerant and local character of the charge carriers that would require multi-band models, and the additional coupling to bosonic modes related to the pseudogap phenomenon. 
\section*{Ackowledgements} 
We thank F. Cilento, G. Coslovich, D. Fausti, F. Parmigiani, D. Mihailovi\'c, P. Prelov\v{s}ek, V.V. Kabanov, U. Bovensiepen, M. Eckstein, A. Avella, D. van der Marel, L. Boeri, L. de' Medici, A. Cavalleri, D. Manske, B. Keimer, D.J. Scalapino for the useful and fruitful discussions. We gratefully acknowledge D. Bonn and B. Keimer for the support in the development of the MPI-UBC Tl2201$_{\mathrm{OD}}$ research effort.
The research activities of S.D.C., F.B, G.F., M.C. and C.G. have received funding from the European Union, Seventh Framework Programme (FP7 2007-2013), under Grant No. 280555 (GO FAST). S.D.C. received financial support by Futuro in Ricerca grant No. RBFR12SW0J of the Italian Ministry of Education, University and Research.
L.V. is supported by the Alexander von Humboldt Foundation.
M.M. acknowledges support from the NCN project DEC-2013/09/B/ST3/01659.
The Y-Bi2212$_{\mathrm{UD}}$ crystal growth work was performed in M.G.Õs prior laboratory at Stanford University, Stanford, CA 94305, USA, and supported by the US Department of Energy, Office of Basic Energy Sciences.
The work at UBC was supported by the Max Planck - UBC Centre for Quantum Materials, the Killam, Alfred P. Sloan, Alexander von Humboldt, and NSERC's Steacie Memorial Fellowships (A.D.), the Canada Research Chairs Program (A.D.), NSERC, CFI, and CIFAR Quantum Materials.
M.C. is financed by European Research Council through FP7/ERC Starting Grant SUPERBAD, Grant Agreement 240524.
J.B. acknowledges support by the P1-0044 of ARRS, Slovenia.
G.C. acknowledges support by the EC under Graphene Flagship (contract no. CNECT-ICT-604391).
N.D.Z. acknowledges support from the NCCR project Materials with Novel Electronic Properties and appreciates expert collaboration with J. Karpinski.

\section*{Author contributions} C.G., S.D.C. and G.C. conceived the experiments. L.V., M.M., J.B. conceived the out-of-equilibrium calculations. C.G. coordinated the research activities with input from all the coauthors, in particular S.D.C., L.V., J.B., M.M., M.C., A.D. and G.C. The broadband pump-probe setup at Politecnico (Milan) was designed and developed by D.B. and G.C. The time-resolved optical measurements were performed by S.D.C., G.S., S.P., L.C., D.B., G.C. The analysis of the time-resolved data have been performed by S.D.C. and C.G. The out-of-equilibrium t-J model calculations have been performed by L.V., D.G., M.M. and J.B. The DMFT calculations were carried out by M.C. The Y-Bi2212 crystals were grown and characterized by H.E., M.G., R.C. and A.D. The Tl-2201 crystals were grown and characterized by L.C. The characterization and the measurement of the equilibrium optical properties of the Bi2201 crystals have been performed by S.L. The MgB$_2$ crystals were grown by N.Z and characterized by B.L. and N.Z. The text was written by C.G. with major input from S.D.C., L.V., M.M., J.B., F.B., G.F., M.G., S.L., M.G., M.C., A.D., D.B. and G.C. All authors extensively discussed the results and the interpretation and revised the manuscript.

\section*{Correspondence} Correspondence and requests for materials
should be addressed to C.G.~(email: claudio.giannetti@unicatt.it), S.D.C.~(email: stefano.dalconte@polimi.it), L.V.~(email: lev.vidmar@lmu.de).

\section*{Methods}
\subsection{Experimental setup}
A Ti:sapphire amplifier (Clark-MXR model CPA-1) delivers a train of pulses at 1 kHz repetition rate with 150-fs duration at 780 nm central wavelength and is used to simultaneously drive three Non-collinear Optical Parametric Amplifiers (NOPAs) operating in different frequency intervals. All NOPAs are seeded by white light continuum (WLC) generated in a sapphire plate. The first NOPA (NOPA1), is pumped by the fundamental wavelength and is amplified by fulfilling quasi-phase matching condition in a periodically poled stoichiometric LiTaO$_3$ (PPSLT) crystal. The signal generated in this process covers a spectral range between 1 $\mu$m (1.24 eV) and 1.5 $\mu$m (0.83 eV) and is temporally compressed to nearly the transform-limited (TL) pulse duration (8.5 fs) by a deformable mirror based pulse shaper \cite{Brida2009}. The second NOPA (NOPA2) is pumped by the second harmonic and amplifies in a beta-barium borate (BBO) crystal pulses with a spectral content between 820 nm (1.5 eV) and 1050 nm (1.2 eV), which are compressed to nearly TL 13-fs duration by a couple of fused silica prisms. Both these NOPAs serve to probe the transient response of the sample and are synchronized with a third NOPA  (NOPA3), pumped by the second harmonic and using BBO, which initiates the dynamics. The spectrum of NOPA3 spans a frequency range between 510 nm (2.4 eV) and 700 nm (1.8 eV) and it is compressed to 7 fs duration by multiple bounces on a pair of chirped mirrors, making the overall temporal resolution of the pump-probe setup below 19 fs. The time delay between pump and probe is adjusted by a motorized delay stage and both the beams are focused on the sample by a spherical mirror in a quasi-collinear geometry. The spectrum of the NOPA1 and NOPA2 probes is detected respectively by an InGaAs and a Si spectrometer working at the full 1 kHz laser repetition rate. By recording the reflected probe spectrum at different temporal delays t with and without pump excitation, we measure the differential reflectivity: $\delta R(\omega,t)/R_{eq}(\omega)$=$[R(\omega,t)$-$R_{eq}(\omega)]/R_{eq}(\omega)$. 

\subsection{Samples}
The Y-substituted Bi2212 single crystals were grown \cite{Eisaki2004} in an image furnace by the travelling solvent floating-zone technique with a non-zero Y content to
maximize $T_{\mathrm{c}}$. The underdoped samples were annealed at 550 $^{\circ}$C for 12 days in a
vacuum-sealed glass ampoule with copper metal inside. La-Bi2201 crystals were all by the floating-zone technique, and characterized as described elsewhere \cite{Ono2003}. Tl2201$_{\mathrm{OD}}$ single crystals were grown \cite{Peets2007} by a copper-rich self-flux method, with stoichiometry Tl$_{1.88(1)}$Ba$_{2}$Cu$_{1.11(2)}$O$_{6+\delta}$ corresponding to Cu substitution on the Tl site, thus away from the CuO$_2$ planes. High-quality MgB$_2$ single crystals were grown using high-pressure cubic anvil technique \cite{Zhigadlo2010}.

\subsection{$t$-$J$ model out of equilibrium}

We investigate the ultrafast energy exchange between doped fermionic carriers (holes) and short-range antiferromagnetic fluctuations within the $t$-$J$ model on a square lattice,
\begin{equation}
H_{tJ} = -t_h \sum_{\langle {\bf ij}\rangle,\sigma}(\tilde c^\dagger_{{\bf i},\sigma} \tilde c_{{\bf j},\sigma} e^{i\phi_{\bf i,j}(\mathrm{t})} +\mathrm{H.c.}) +
\sum_{\langle {\bf ij}\rangle } J ( {\bf S}_{\bf i} \cdot {\bf S}_{\bf j} - \frac{1}{4}\tilde{n}_{\bf i} \tilde{n}_{\bf j} ),
\label{def_tj}
\end{equation}
where $\tilde c_{{\bf i},\sigma}$=$c_{{\bf i},\sigma}(1 - n_{{\bf i},-\sigma})$ is a projected fermion operator, $t_h$ represents nearest neighbour overlap integral, the sum $\langle \bf ij \rangle$  runs over pairs of nearest neighbours and
$\tilde n_{\bf i}$=$n_{\bf{i},\uparrow}$+$n_{\bf{i},\downarrow}$-$2 n_{\bf{i},\uparrow} n_{\bf{i},\downarrow}$ is a projected electron number operator. The system is threaded by a time-dependent flux $\phi_{\bf i,j}$(t), which induces the electric field $ - \partial_t {\phi}_{\bf i,j}$(t).

Motivated by the experimental results, we model the dynamics far from equilibrium under the following assumptions:
(i)
We mimic the effect of an ultrashort laser pump pulse by instantly increasing the kinetic energy of a doped carrier, while the chargeless spin degrees of freedom remain unchanged.
In the context of the $t$-$J$ model in Eq.~(\ref{def_tj}), we first calculate the ground state with $\phi_{\bf i,j}$=$0$, then we perform a sudden change (quench) of the phase at time t=0, describing the $\delta$-like pulse of the electric field.
Such a phase quench is obtained by setting  $\phi_{\bf i,i+e_{x(y)}}$(t)=$\pi \theta(t)$, where $\bf e_{x(y)}$ represents the lattice vector in the  $x(y)$-direction.
(ii)
In a generic situation, promptly after the absorption of the pump  pulse, the kinetic energy of only a fraction of holes (also referred to as photo-excited holes) is suddenly increased.
For optimally doped samples ($p\sim$0.16) and the pump pulse with photon energy $2$~${\rm eV}$, fluence $700$~${\rm \mu J/cm^2}$ and penetration depth $170$~${\rm nm}$, the doping rate representing photo-excited holes is rather low, $p_0\sim0.01$, assuming that each photo-excited hole absorbs the entire quantum of light, {\it i.e.}, $2$~${\rm eV}$.
We therefore study the dynamics of a single photo-excited hole, assuming that the main relaxation mechanism is represented by coupling of the photo-excited hole to local antiferromagnetic excitations.
As a consequence of finite chemical and photo doping, the average number of antiferromagnetic bonds, which are available to absorb the excess kinetic energy of a single photo-excited hole, is large but finite.  
While our method is limited to the calculation of single hole dynamics, we simulate the effect of small photo doping by limiting the propagation of the photo-excited hole to a finite $L\times L$ plaquette with $L<10$.

Since exact diagonalization (ED) is limited to clusters which are too small to account for the complete relaxation within the experimental conditions described above,
we extend the lattice size by employing diagonalization in a limited functional space.
The latter method was recently successfully applied to describe non-equilibrium properties of a charge carrier propagating in a local antiferromagnetic background \cite{mierzejewski2011,Golez2013}. 
The advantage of this method over the standard ED in the equilibrium regime follows from a systematic construction of states with distinct configurations of local antiferromagnetic excitations in the proximity of the hole. In addition, it remains efficient even when applied to non-equilibrium systems on short and intermediate time scales, as long as the antiferromagnetic disturbance caused by the local quench remains within the boundaries of generated excitations \cite{mierzejewski2011,Golez2013}.
We first construct the parent state
\begin{equation}
 |\varphi^{(0)}\rangle=c_{\mathbf{k},\sigma}|\mbox{N\' eel}\rangle, \label{wavef_start}
\end{equation}
which represents a hole with momentum $\mathbf{k}$, doped into the N\' eel state.
The functional space \cite{Bonca2007} is then generated by
\begin{equation} \label{def_lfs}
 \left\{|\varphi_{j}^{(n_h)} \rangle \right\} = \left[ H_{\rm kin} + \tilde H_J \right]^{n_h} |\varphi^{(0)} \rangle, \; \; n_h=0,...,N_h,
\end{equation}
where $H_{\rm kin}$ represent the kinetic energy term of the $t$-$J$ model and $\tilde H_J$ represents the spin-flip term of the Heisenberg part of the $t$-$J$ model (spin-flip denotes overturned spins with respect to the N\' eel state; $\tilde H_J$ erases two neighboring spin-flips created by the propagating hole).
The size of the functional space is determined by the parameter $N_h$.
In addition, we impose the condition that the largest distance of a local antiferromagnetic excitation away from the photo-excited hole, $\tilde{\bf r}$=$(N_{\rm box},N_{\rm box})$, should not exceed $N_{\rm box}=4$, therefore limiting the spatial extent of antiferromagnetic excitations to a $9$$\times$$9$ plaquette.
Contrary to the standard ED, we carry out a finite-size scaling not with respect to the geometric size of the system but rather with respect to $N_h$ (see Supplementary Information).

Applying the Lanczos technique we first compute the ground state $|\Psi(\mathrm{t}$=$0)\rangle$.
The time evolution $|\Psi(\mathrm{t})\rangle$=$e^{-iH_{tJ}\mathrm{t}} |\Psi(\mathrm{t}$=$0)\rangle$ is then implemented using the quenched Hamiltonian.
We set the hopping amplitude $t_h$=$0.360$~eV, which corresponds to the time unit $\hbar/t_h$=$1.83$~fs.
At each time step $\delta \mathrm{t} (t_h /\hbar)$$\ll$1 we generate the evolution $|\Psi(\mathrm{t}$-$\delta \mathrm{t})\rangle$$\rightarrow$$|\Psi(\mathrm{t})\rangle$ by using the Lanczos basis \cite{park1986}.
We measure in Figure 2a the time-dependent expectation value of the kinetic energy relative to the initial state, $\Delta E_{\rm kin}(\mathrm{t})$=$E_{\rm kin}(\mathrm{t})$-$E_{\rm kin}^{(0)}$, where $E_{\rm kin}(\mathrm{t})$=$\langle H_\mathrm{kin}(\mathrm{t})\rangle$ and $E_{\rm kin}^{(0)}$=$\langle H_\mathrm{kin}(\mathrm{t}$$<$$0)\rangle$.
In Figure 2b we plot the relative increase of the spin energy on the antiferromagnetic bond at a distance ${\bf r}$ away from the photo-excited hole, $\Delta E_{\rm s}(\mathrm{t},{\bf r})$=$E_{\rm s}(\mathrm{t},{\bf r})$-$E_{\rm s}^{(0)}({\bf r})$, where $E_{\rm s}(\mathrm{t},{\bf r})$=$J\langle {\bf S}_{\bf i+r}(\mathrm{t})$$\cdot$${\bf S}_{\bf j+r}(\mathrm{t}) \rangle$ and $E_{\rm s}^{(0)}({\bf r})$=$J\langle {\bf S}_{\bf i+r}(\mathrm{t}$$<$$0)$$\cdot$${\bf S}_{\bf j+r}(\mathrm{t}$$<$$0) \rangle$.
After the quench, the total energy in the system remains constant, $\Delta E_{\rm kin}(\mathrm{t})$+$\sum_{\bf r}\Delta E_{\rm s}(\mathrm{t},{\bf r})$=${\rm const.}$



\clearpage
\newpage


\begin{titlepage}
\begin{center}

\textsc{\Large Supplementary Material for:\textit{Snapshots of the retarded interaction of charge carriers with ultrafast fluctuations in cuprates} }\\[2.5cm]

\text{\normalsize Contents}\\[1.5cm]

\end{center}
\text{\normalsize I. SVD ANALYSIS}\\[0.5cm]
\text{\normalsize II. TRANSIENT REFLECTIVITY MEASUREMENTS AT LOW TEMPERATURE}\\[0.5cm]
\text{\normalsize III. EXTENDED DRUDE MODEL AND SCATTERING RATE}\\[0.2cm]
\text{\normalsize A. Equilibrium optical response}\\[0.2cm]
\text{\normalsize B. Non-equilibrium optical response and the isosbestic point}\\[0.2cm]
\text{\normalsize C. Magnesium diboride}\\[0.5cm]
\text{\normalsize IV. TEMPORAL RESOLUTION}\\[0.5cm]
\text{\normalsize V. RELAXATION DYNAMICS IN THE t-J MODEL}\\[0.2cm]
\text{\normalsize A. Finite-size scaling}\\[0.2cm]
\text{\normalsize B. The role of a direct carrier-carrier interaction in the lower Hubbard band}\\[0.5cm]
\text{\normalsize VI. DYNAMICAL MEAN-FIELD THEORY (DMFT)}\\[0.5cm]
\text{\normalsize References}\\[0.5cm]

\end{titlepage}



\clearpage
\newpage

\setcounter{figure}{0}
\setcounter{equation}{0}

\makeatletter 
\renewcommand{\thefigure}{S\@arabic\c@figure}
\makeatother

\makeatletter 
\renewcommand{\theequation}{S\@arabic\c@equation}
\makeatother

\section{SVD analysis}
The Y-Bi2212$_\mathrm{UD}$ non-equilibrium reflectivity $\frac{\delta R(\omega,t)}{R_{eq}(\omega)}$, reported in Fig.~1b of the main text, is obtained by combining the transient response measured in two spectral windows with laser pulses provided by NOPA1 and NOPA2 (see Methods). In general, the $\frac{\delta R(\omega,t)}{R_{eq}(\omega)}$ matrix contains different spectral features and temporal dynamics, which are characteristic of the physical processes triggered by the pump beam. In order to analyze the 2D $\frac{\delta R(\omega,t)}{R_{eq}(\omega)}$ matrix, different approaches are feasible. For example, one can model the reflectivity variation at a fixed delay time $\mathrm{\tilde{t}}$, i.e., $\frac{\delta R(\omega,\tilde{t})}{R_{eq}(\omega)}$, and describe the entire 2D map as the result of the temporal variation of some parameters used to describe the spectral response S\cite{Giannetti2011}. However, this approach turns out to be difficult to apply when the response of the sample cannot be described as a single component. In order to determine how many components contribute to the full $\frac{\delta R(\omega,t)}{R_{eq}(\omega)}$ matrix, we adopt a method based on the singular value decomposition (SVD) analysis. This method has been previously used to study the interplay between the non-equilibrium dynamics of the pseudogap and the superconducting phases in YBi2212 S\cite{Coslovich2013}. It can be shown that a generic $m\times n$ matrix, M, can be factorized in the form $M=TSV$ where T is a $m\times m$ unitary matrix, S is a non negative $m\times n$ diagonal matrix which contains the singular values (in decreasing order with increasing row number) and V is a $n\times n$ matrix. This factorization is such that the $j^{th}$ column of T and the $j^{th}$ row of V contain, respectively, the trace in the temporal and spectral domains corresponding to the $j^{th}$ singular value of the S matrix. The SVD LAPACK routine is used to compute the decomposition and to extract the principal components (singular values) which contribute to the overall optical response. It is worth to stress that, since the eigenvalues of the S matrix are non-degenerate, the SVD decomposition is unique i.e. it is always possible to find for each singular value a unique couple of temporal and spectral eigenvectors.
Using SVD, the experimental $\frac{\delta R(\omega,t)}{R_{eq}(\omega)}$ matrix can be written as:

\begin{equation}\label{factoriz}
\frac{\delta R (\omega,\mathrm{t})}{R_{eq}(\omega)}=\sum_{j}k_{j}\Psi_{j}(\omega)\otimes\delta\phi_{j}(t)
\end{equation}

\begin{figure}[b]
	\centering
		\includegraphics[keepaspectratio,clip,width=0.9\textwidth] {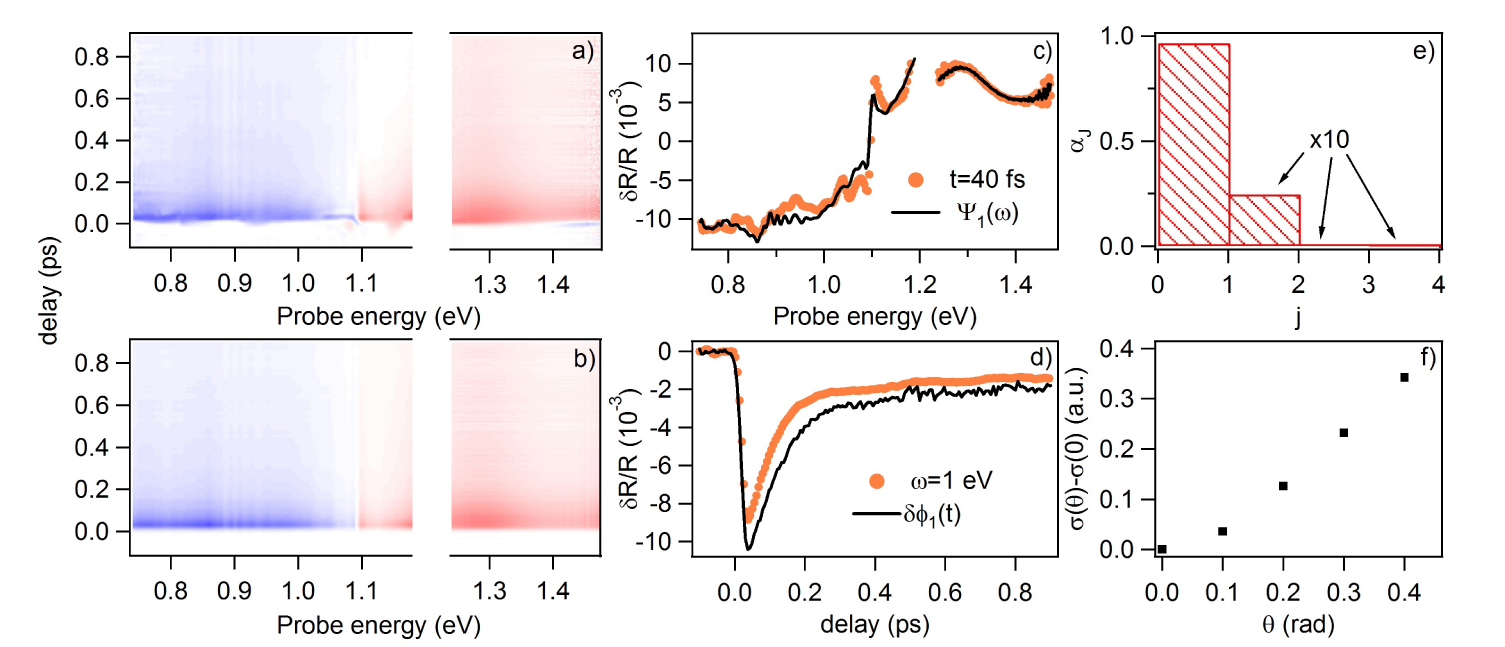}
	\caption{\textbf{a) Transient reflectivity variation of Y-Bi2212$_\mathrm{UD}$. b) Time-frequency matrix calculated as the outer product between $\Psi_{1}(\omega)$ and $\delta\phi_{1}(t)$ multiplied by the first singular value $k_{1}$ c) Energy trace at t=40~fs, corresponding to the maximum variation of the transient signal, (orange dots) and the first spectral eigenvector (black line) $\Psi_{1}(\omega)$ multiplied by $\delta\phi_{1}$(t=40~fs) and $k_{1}$ d) Time trace at $\hbar\omega$=1~eV displayed with the first temporal eigenvector $\delta\phi_{1}$(t) weighted by $\Psi_{1}$($\omega$=1~eV) and $k_{1}$. For graphical reasons, $\delta\phi_{1}$(t) has been offset. e) Weights $\alpha_{j}$ of the singular values normalized to one ($\alpha_{j}=\frac{k^{2}_{j}}{\sum_{j}k^{2}_{j}}$) and sorted in decreasing order f) $\sigma(\theta)$, defined in Eq.~\ref{sigma_parameter}, computed for different value of the rotation angle $\theta$.}}
\label{Figura1}
\end{figure}

where $k_{j}$ is the $j^{th}$ singular value while $\Psi_{j}(\omega)$ and $\delta\phi_{j}(t)$ are the corresponding $j^{th}$ spectral and temporal eigenvectors. The results of the SVD analysis on the Y-Bi2212$_\mathrm{UD}$ spectral and temporal response are reported in Fig.~\ref{Figura1}. The main outcome of this analysis is that the first eigenvalue alone accounts for more than 97~$\%$ of the global response. This result can be appreciated by comparing the raw data with the matrix $\frac{\delta R}{R}_{1}$ built as the outer product between the first temporal and spectral eigenvectors, weighted by the first singular value ($\frac{\delta R}{R}_{1}=k_{1}\Psi_{1}(\omega)\otimes\delta\phi_{1}(t)$), reported respectively in Fig.~\ref{Figura1}a and b. Both the temporal and spectral behavior is perfectly reproduced by the first temporal and energy eigenvector as shown in Fig.~\ref{Figura1}c and d. The weights of the singular values, normalized to the unity, are reported in Fig.~\ref{Figura1}e: the weight of the eigenvalues $\alpha_{j}$, with $j>1$, are multiplied by ten to emphasize the difference with respect to the first one. In order to highlight the uniqueness of the SVD factorization and the effectiveness of the approximation with only the first component, the matrix of the temporal and spectral traces are rotated by the same angle $\theta$ and a new matrix $\frac{\delta R}{R}_{1}(\theta)$ is computed as the outer product of the rotated vectors: $\frac{\delta R}{R}_{1}(\theta)=k_{1}R(\theta)\Psi_{1}(\omega)\otimes\delta\phi_{1}(t)R(\theta)$ where $R(\theta)$ is the rotation matrix. The mean-square deviation between the experimental data $\frac{\delta R(\omega,t)}{R(\omega)}$ and the computed matrix $\frac{\delta R}{R}_{1}(\theta)$ is:

\begin{equation}\label{sigma_parameter}
\sigma(\theta)=\frac{\sqrt{\sum_{i,j}\left(\frac{\delta R(\omega,t)}{R(\omega)}_{i,j}-\frac{\delta R}{R}_{1i,j}\right)^{2}}}{\sqrt{\sum_{i,j}\left(\frac{\delta R(\omega,t)}{R(\omega)}_{i,j}\right)^{2}}}
\end{equation}

and it is reported in Fig.~\ref{Figura1}f for different values of $\theta$. $\sigma(\theta)$ displays a minimum for $\theta$=0 and it increases for any kind of rotation of the eigenvectors, indicating that $\frac{\delta R}{R}_{1}$ is the best approximation of the experimental matrix.



\section{Transient reflectivity measurements at low temperature}
\begin{figure}[b]
	\centering
		\includegraphics[keepaspectratio,clip,width=0.5\textwidth] {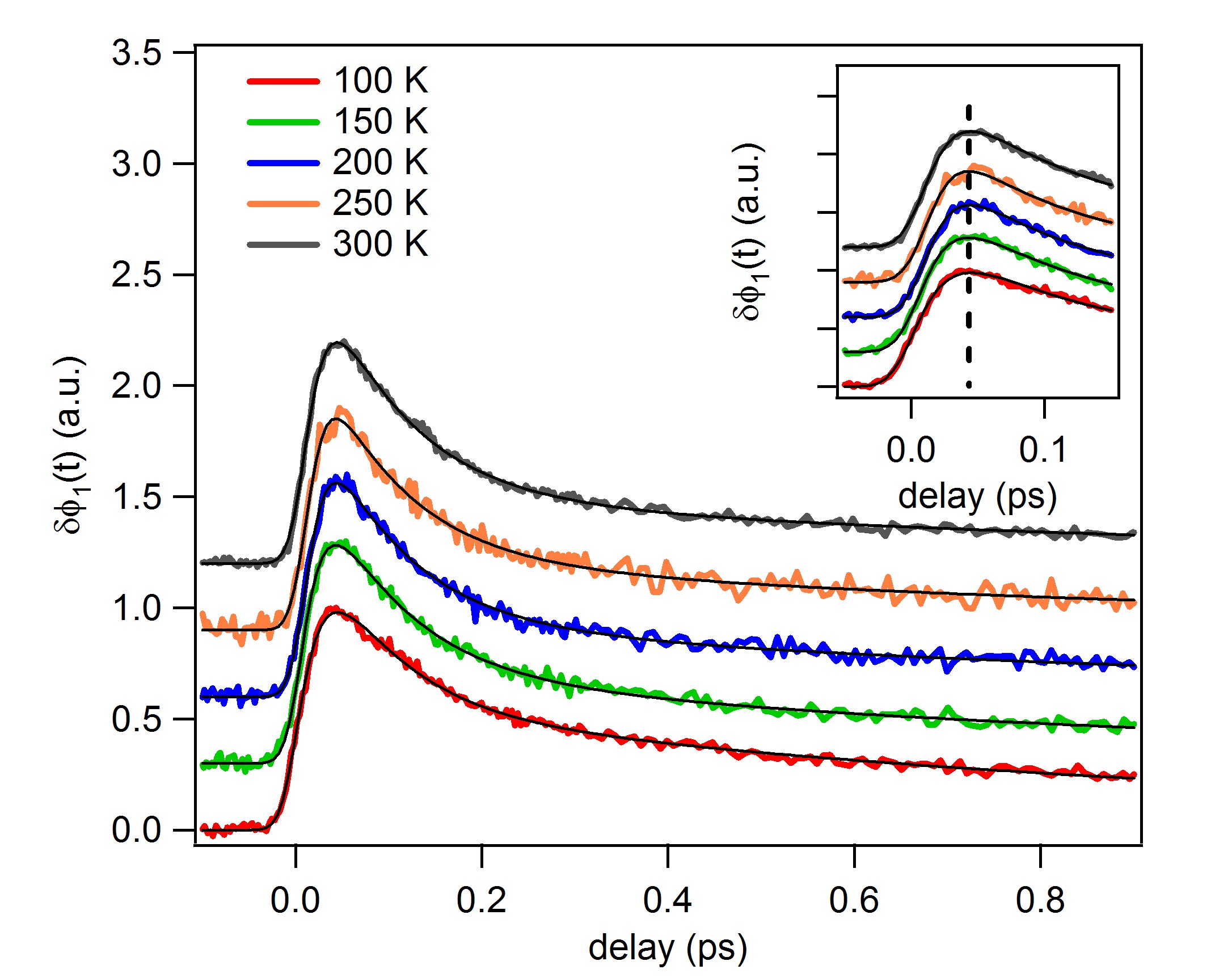}
	\caption{\textbf{Temporal dynamics of $\delta\phi_{1}$(t) eigenvectors at different temperatures. The black lines are the fit to the data. The fitting function takes into account the experimental resolution, as described more extensively in the main text. In the inset, the dynamics at short delay times clearly show that the rise time of the temporal traces does not depend on the temperature.}}
\label{Figura2}
\end{figure}
Transient reflectivity measurements on Y-Bi2212$_\mathrm{UD}$ single crystal are carried out at low temperature in a liquid nitrogen cryostat. The minimum temperature achievable in the cryostat is about 100~K. The temperature of the sample is monitored by a thermocouple placed on the sample holder. The high temporal resolution (below 20~fs) is preserved by using a very thin (200~$\mu$m) UV fused silica small flange as optical access to the cryostat. The first temporal eigenvectors $\delta\phi_{1}(t)$ extracted by the SVD analysis of the experimental matrices at different temperatures, are reported in Fig.~\ref{Figura2}. All the time traces do not display any change in the dynamics in the 100~K-300~K temperature range reaching the maximum variation at the same delay time ($\approx$40~fs) as shown in the inset. 

\section{Extended Drude model and scattering rate}


\subsection{Equilibrium optical response}
\begin{figure}[t]
	\centering
		\includegraphics[keepaspectratio,clip,width=0.5\textwidth] {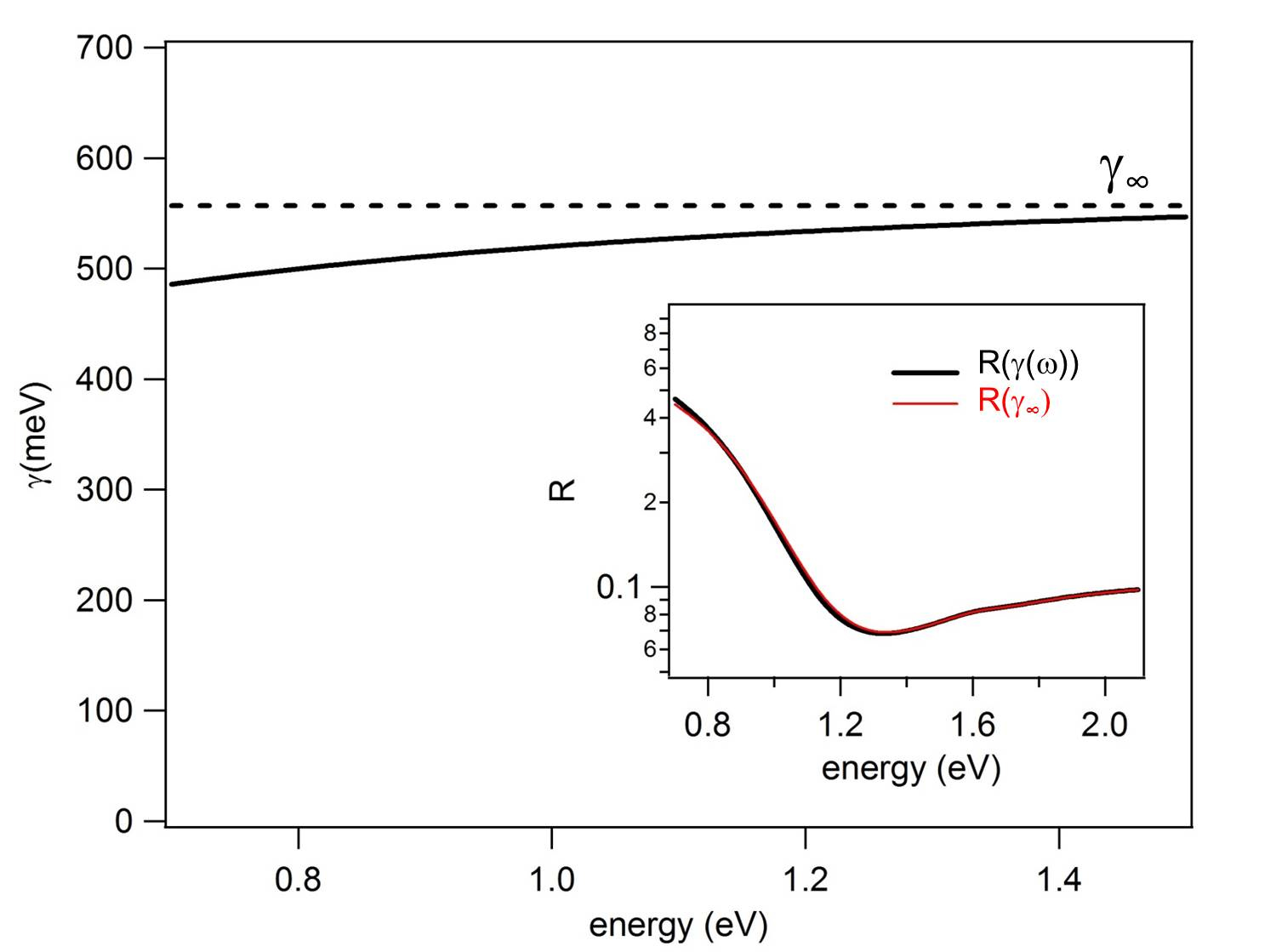}
	\caption{\textbf{Energy dependence of the scattering rate at T=300~K in the probe spectral range. $\gamma_{\infty}$ is the asymptotic value of the scattering rate. In the inset the EDM fit to static reflectivity (black line) is compared with the calculated reflectivity (red line) under the frequency independent approximation of the damping term: $\gamma(\omega)=\gamma_{\infty}$.}}
\label{Figura5}
\end{figure}

The equilibrium reflectivity $R_{eq}(\omega)$, measured by spectroscopic ellipsometry S\cite{DalConte2012}, is reported in Fig.~\ref{fig1}a (continuum line) of the main text for the Y-Bi2212$_\mathrm{UD}$ sample (hole doping p=0.13, T$_{c}$=83~K). $R_{eq}(\omega)$ is related to the complex equilibrium dielectric function $\epsilon(\omega)$ by:

\begin{equation}\label{Refl}
R_{eq}(\omega)=\left|\frac {1-\sqrt{\epsilon(\omega)}}{1+\sqrt{\epsilon(\omega)}}\right|^{2}
\end{equation}
where $\epsilon(\omega)=1+i\frac {4\pi\sigma(\omega)}{\omega}$.
A model dielectric function, $\sigma(\omega)=\sigma_{D}(\omega)+\sum_{i}\sigma_{L_{i}}(\omega)$, which combines the Extended Drude term (see main text) and a sum of Lorentz oscillators centered at energies $\approx$1.5, 2, 2.7~eV ($\sigma_{L_{i}}(\omega)$), that account for the interband transitions in the visible range, is used to reproduce the experimental dielectric function over the 0-3.5~eV energy range. 
The bare plasma frequency resulting from the fit is $\omega_{p}$=2.3~eV, in agreement with the values from the literature for similar systems close to the optimally doped level S\cite{VanHeumen2009}. 

A direct expression of the Extended Drude Model (EDM) scattering rate, $\gamma(\omega,T)$, in terms of the electron-boson coupling $I^{2}\chi(\Omega)$ can be derived from the Kubo formula of the optical conductivity in the framework of the electron-boson Holstein theory S\cite{Allen1971}. In this case, the electron self energy, $\Sigma(\omega,T)$, can be calculated as a convolution integral between $I^2$$\chi(\Omega)$ and a kernel function $L(\omega,\Omega,T)$:
\begin{equation}
\label{selfenergy}
\Sigma(\omega,T)=\int_0^\infty I^2 \chi(\Omega)L(\omega,\Omega,T)d\Omega
\end{equation}
where the kernel function can be calculated analytically and it has the following expression:
\begin{equation}
\label{kernel}
L(\omega,\Omega;T)=-2\pi i\left[n(\Omega,T)+\frac{1}{2}\right] +\Psi \left(\frac{1}{2}+i\frac{\Omega-\omega}{2\pi T}\right) -\Psi \left(\frac{1}{2}-i\frac{\Omega+\omega}{2\pi T}\right)
\end{equation}
It is worth to note that the kernel function describes the thermal excitation of both the bosonic modes through the Bose-Einstein distribution $n(\Omega,T)$, and the electronic carriers through the terms containing the digamma functions $\Psi$. Consequently, in the first term of eq.~\ref{kernel}, $T$ is the temperature of the bosonic excitations while, in the rest of the formula, is the temperature of the electronic carriers.   
For moderately doped systems, in which the vertex corrections can be omitted, the optical scattering rate is related to the single-particle self-energy by:
\begin{equation}
\label{scatteringrate}
\gamma(\omega,T)=\mathrm{Im}\left\{\omega\left[\int_{-\infty}^{+\infty} \frac{f(\xi,T)- f(\xi+\omega,T)}{\omega+\Sigma^*(\xi,T)-\Sigma(\xi+\omega,T)} d\xi \right]^{-1}-\omega\right\}
\end{equation}
where $f$ is the Fermi-Dirac distribution, $\Sigma(\omega,T)$ and $\Sigma^*(\omega,T)$ the electron and hole \textbf{k}-space averaged self-energies. 

The bosonic glue, extracted from the fit to the equilibrium dielectric function, extends up to $\approx$300~meV, and displays a narrow peak at $\approx$70~meV. A more detailed description of the $I^{2}\chi(\Omega)$ spectrum can be found in Refs.~S\cite{VanHeumen2009,DalConte2012,Carbotte2011}. The scattering rate calculated from Eq.~\ref{scatteringrate} is reported in Fig.~\ref{Figura5} and displays a variation of about 10~$\%$ in the entire probe energy window. Because of the weak energy dependency of $\gamma(\omega)$, in the 0.7-1.5~eV energy range, the equilibrium reflectivity can be reasonably described in this spectral region by the asymptotic value $\gamma_{\infty}$ as it is shown in the inset of Fig.~\ref{Figura5}.

\subsection{Non-equilibrium optical response and the isosbestic point}
\begin{figure}[b]
	\centering
		\includegraphics[keepaspectratio,clip,width=0.8\textwidth] {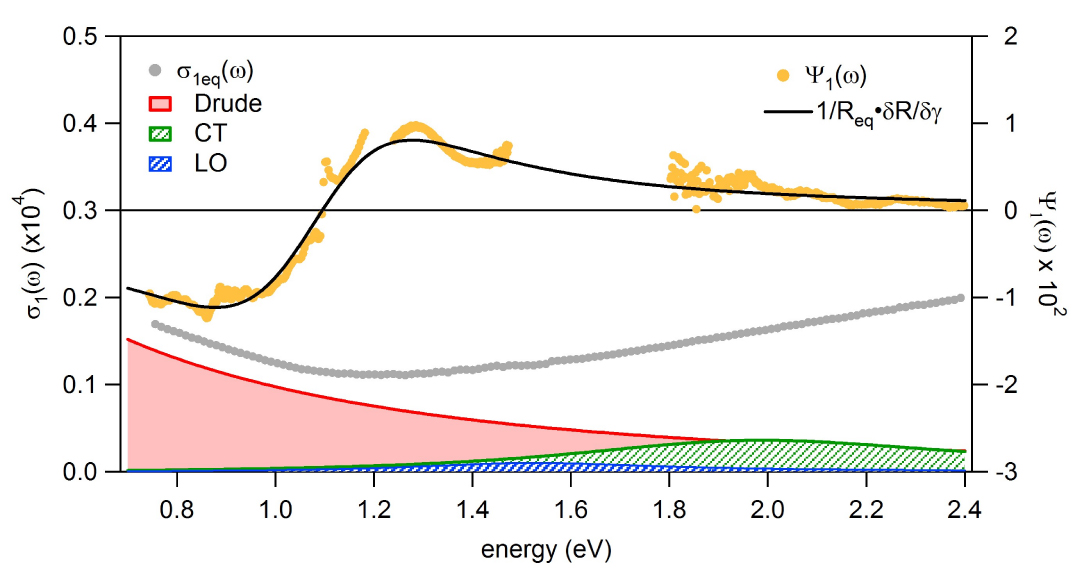}
	\caption{\textbf{The experimental equilibrium optical conductivity (grey dots) together with the Drude and charge transfer peak resulting from the fit of the equilibrium optical conductivity (red and the green lines). The first eigenvalue obtained from the SVD analysis is reported (yellow dots). The trace extending between 0.7 and 1.5~eV is the one previously reported in Fig.~1 of the main text. The trace extending between 1.8~eV and 2.4~eV results from the SVD analysis of the Y-Bi2212$_\mathrm{UD}$ transient response probed by NOPA 3. The black line is the best fit obtained by increasing the total scattering rate.}}
\label{Figura7}
\end{figure}

In time-resolved experiments, the pump pulse drives the system out of equilibrium while a second broadband probe pulse monitors the transient change of the dielectric function. In general, the relaxation dynamics can follow two different relaxation pathways. In the first scenario, the pump energy is absorbed by a small fraction of the charge carriers which scatter with the other unexcited carriers and relax on a very short timescale towards a Fermi-Dirac distribution characterized by a temperature higher than that the equilibrium. One can rationalize this process in term of a sudden increase of the electronic temperature ($\delta T_{e}>$0) while the bosonic bath is essentially unchanged ($\delta T_{b}=$0). In the second scenario, the excited carriers mainly release the excess energy through scattering with the bosonic modes. In terms of the effective temperature models, this picture implies that $\delta T_{e}\simeq\delta T_{b}$ at all the timescales longer than the single electron-boson scattering process. The rapid increase of the bosonic excitations results, in turn, in the increase of the optical scattering rate ($\delta\gamma_{\infty}>$0), which can be detected under the form of a broadening of the Drude peak. 

As it was already demonstrated (see Fig.~1 in S\cite{DalConte2012}), the two physical scenarios previously sketched, result in a dramatically different reflectivity variation in the frequency domain. In Fig. 1 of the manuscript we demonstrate that the experimental $\delta R(\omega$,t) is reproduced by assuming an increase of $\delta\gamma_{\infty}$ on a extremely short time scale ($<$50~fs). No signature of the electronic carriers decoupled from the bosonic degrees of freedom has been revealed. These results are compatible only with the second scenario, in which the charge carriers mainly exchange energy with the bosonic bath. Therefore, we can assume that $\delta R$(t)$\propto\delta\gamma_{\infty}$(t).

Having clarified how the transient reflectivity reveals the dynamics of the energy stored in the bosonic bath, we focus now on the non-equilibrium optical response around the isosbestic point. 
In particular, it can be shown that all the reflectivity curves corresponding to different scattering rates cross each other at a precise frequency $\tilde{\omega}\approx$1.1~eV, which is called isosbestic point. The physical properties of the isosbestic points are discussed in a recent work S\cite{Greger2013}. Across an isosbestic point a physical quantity $f(\omega,g)$ can be locally approximated by an expansion of a generic parameter $g$ around some particular value $g_{0}$. In this specific case, for a small variation of the scattering rate, the following expansion of the reflectivity holds:

\begin{equation}\label{expa}
R(\omega,\gamma)=R(\omega,\gamma_{0})+\frac{\partial R}{\partial \gamma}(\gamma-\gamma_{0})+ O\left[(\gamma-\gamma_{0})^{2}\right]
\end{equation}

The effect of the pump pulse, in a pump-probe experiment, is to transiently perturb the optical properties of the system. In Fig.~1 of the main text we show that $\delta R$($\omega$,$\mathrm{\tilde{t}}$)/$R_{eq}(\omega)$ changes sign from negative to positive at the frequency corresponding to $\tilde{\omega}$. The major point of the analysis is that the build up dynamics of $\delta R$($\omega$,$\mathrm{\tilde{t}}$)/$R_{eq}(\omega)$, immediately after the pump excitation, can be reproduced solely by assuming a transient increase of the electron-boson scattering rate. This result can be quantitatively expressed in terms of the variation of $\gamma_{\infty}$. The maximum transient signal, occurring at a delay time of $\approx$40~fs for Y-Bi2212$_\mathrm{UD}$, is equivalent to a relative variation $\frac{\delta\gamma_{\infty}}{\gamma_{\infty}}\approx$1~$\%$. Therefore Eq.~\ref{expa} can be reformulated in this way: 

\begin{equation}\label{expa02}
\frac{\delta R(\omega,\mathrm{t})}{R_{eq}(\omega)}\approx\frac{1}{R_{eq}(\omega)}\frac{\partial R(\omega)}{\partial\gamma_{\infty}}\delta\gamma_{\infty}(t).
\end{equation}

Comparing Eqs.~\ref{expa} and \ref{expa02}, we can easily identify the dynamics of the scattering rate variation $\delta\gamma_{\infty}(t)$ with the first temporal eigenvalue obtained through the SVD $\delta\phi_{1}$(t).

To further support the interpretation of the experimental results solely as a change of the scattering rate, the non-linear optical response is measured in the visible-UV range between 1.8~eV to 2.4~eV using chirped-mirrors compressed pulses in a degenerate configuration (both pump and probe pulses are generated by NOPA 3). In this energy region the high energy tail of the Drude peak overlaps with the interband optical transitions. In particular, a weak additional Lorentz oscillator (LO) at $\approx$1.5~eV and a more intense one at $\approx$2~eV are necessary to reproduce the equilibrium dielectric function. The peak at 2~eV is usually interpreted as the reminiscence of the charge transfer (CT) edge in the undoped compound, arising from the excitation of an hole from the upper Hubbard band to the oxygen $2p_{x,y}$ orbital, and it should be particularly sensitive to rearrangements of the oxygen orbitals. Fig.~\ref{Figura7} shows that the first eigenfunction ($\Psi_1(\omega)$) of the SVD of the $\delta R$($\omega$,t)/$R_{eq}(\omega)$ matrix can be reproduced by assuming an ultrafast broadening of the Drude peak, i.e., $\delta\gamma_{\infty}$$>$0. This result rules out any possible variation of the position of the oxygen orbitals during the ultrafast relaxation process, since the transient variation of the parameters of the CT peak would provide a narrow reflectivity variation around 2~eV, incompatible with the measured monotonic variation of the reflectivity.

\subsection{Magnesium diboride}
\begin{figure}[t]
	\centering
		\includegraphics[keepaspectratio,clip,width=1\textwidth] {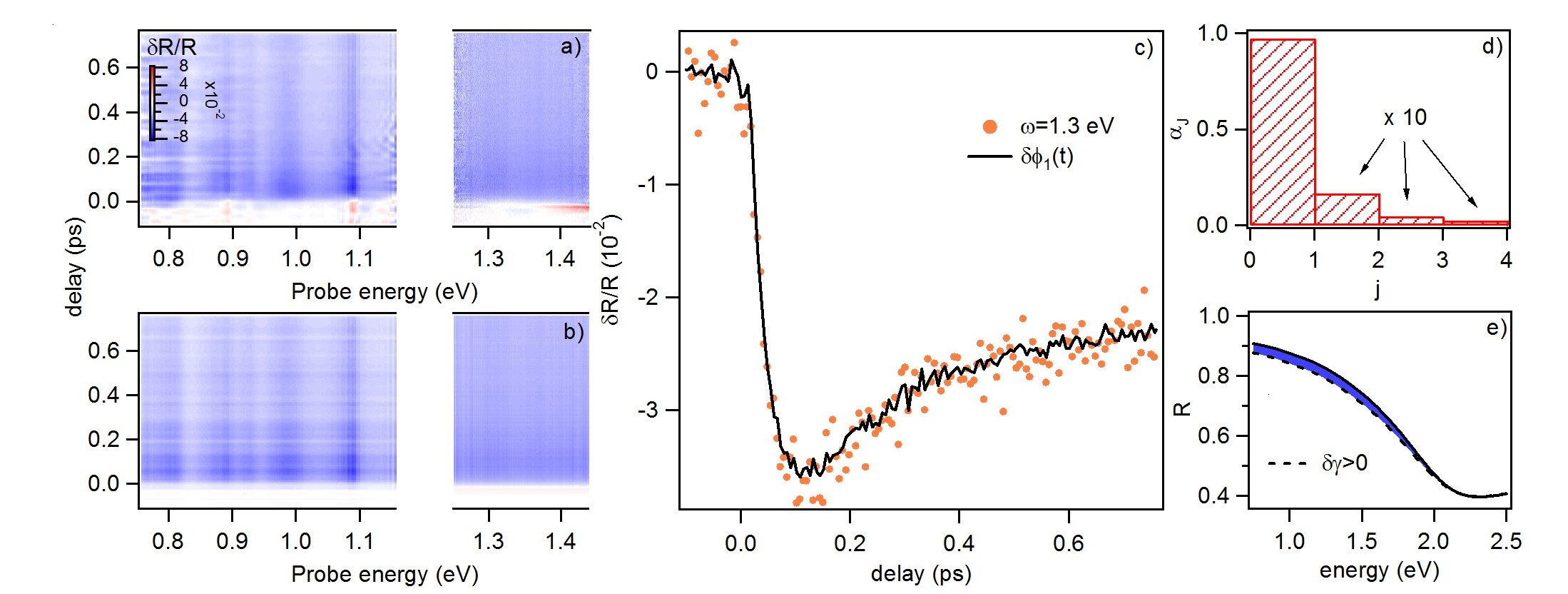}
	\caption{\textbf{a,b) MgB$_{2}$ transient reflectivity variation compared with the matrix calculated as the outer product between $\Psi_1(\omega)$ and $\delta\phi_{1}$(t) multiplied by the first singular value $k_{1}$ c) Time trace at $\hbar\omega$=1.3~eV displayed with the first temporal eigenvector $\delta\phi_{1}$(t) weighted by $\Psi_{1}$($\omega$=1.3~eV) and $k_{1}$ d) Weights $\alpha_{j}$ of the singular values normalized to one ($\alpha_{j}=\frac{k^{2}_{j}}{\sum_{j}k^{2}_{j}}$) and sorted in decreasing order e) MgB$_{2}$ in plane static reflectivity measured at T=300~K (black line) and the static reflectivity calculated within the framework of the EDM by simulating an increase of the scattering rate (dashed line).}}
\label{Figura3}
\end{figure}
The MgB$_{2}$ transient reflectivity is reported in Fig.~\ref{Figura3}a and is factorized by the SVD. Analogously to Y-Bi2212$_\mathrm{UD}$, the MgB$_{2}$ non-equilibrium optical response is entirely described by the first temporal and spectral eigenvectors as shown in Fig.~\ref{Figura3}b. The temporal eigenvector $\delta\phi_{j}$(t) perfectly overlaps with the experimental temporal traces (Fig.~\ref{Figura3}c) proving the reliability of the description of the experimental matrix in term of the highest weight eigenvectors. Despite its high T$_{c}$, it is generally accepted that in MgB$_{2}$, the Cooper pair formation is mediated by optical phonons at 70~meV energy, while electronic correlations are expected to play a minor role. The MgB$_{2}$ equilibrium reflectivity along the $ab$ plane displays a broad plasma edge at a frequency that is blue-shifted ($\approx$2~eV) as compared to Y-Bi2212$_\mathrm{UD}$. Moreover an additional Lorentz peak at $\approx$2.6~eV is also observed in the visible-UV optical range S\cite{Guritanu2006} due to the $\sigma\rightarrow\pi$ transition close to the M point of the Brillouin zone. Transient modifications of this interband transition are expected to occur on an energy scale larger than the probe spectral window and are not expected to influence the non equilibrium response. In Fig.~\ref{Figura3}e the MgB$_{2}$ equilibrium reflectivity (black line) and the reflectivity obtained by increasing the scattering rate (dashed line) in the EDM that reproduces the equilibrium optical data are reported. Also in this case, the transient response, reported in Fig.~\ref{Figura3}b of the main text, can be rationalized as a broadening of the Drude peak, i.e., an increase of the scattering rate.
The $\gamma_{\infty}$(t) curve for MgB$_{2}$ shows a build-up time, $\tau_r$=32$\pm$3~fs, that is significantly larger than that observed on cuprates. This physical process is related to the thermalization with the optical bond stretching phonons at 70~meV, which display a high coupling strength and are considered at the origin of the high T$_{c}$ of MgB$_{2}$ S\cite{Kong2001}.

\section{Temporal resolution}

\begin{figure}[t!]
	\centering
	\includegraphics[keepaspectratio,clip,width=0.8\textwidth] {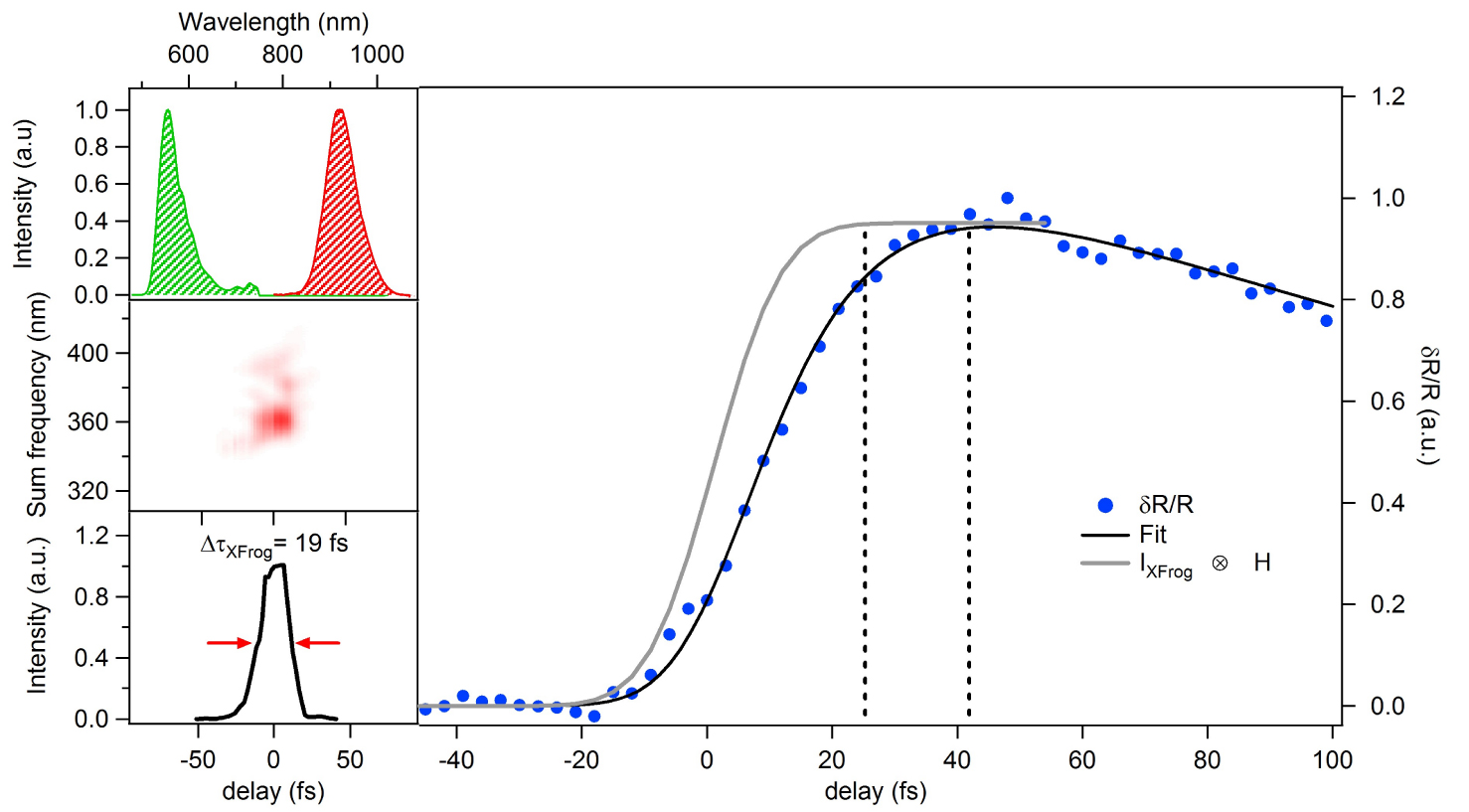}
	\caption{\textbf{a) Spectrum of the pump (OPA3) and the probe (OPA2) pulses; b) Measured XFROG trace; c) Cutoff of the XFROG trace taken at the maximum of the signal (360~nm); d) Transient reflectivity at $\lambda_{probe}$=930~nm (blue dots), fit of the data (black line), convolution between Heaviside function H and the XFROG trace at $I_{XFROG}$ (grey line).}}
\label{Figura6}
\end{figure}

As it was extensively discussed in the manuscript, the build-up dynamics of $\gamma_{\infty}$(t) is strictly related to the time scale on which the charge carriers exchange their excess energy with short range antiferromagnetic excitations. Therefore, the experimental temporal resolution is a crucial parameter to obtain a precise measurements of the intrinsic time scale of this process. To characterize the temporal resolution of the experiment, the pump and the probe pulses are sent through a 20~$\mu$m thick BBO crystal and the spectrum of the sum-frequency generated is detected at different delays. The small thickness of the non linear crystal minimizes the temporal broadening and guarantees a broadband phase matching for all the spectral components of the two pulses. The cross-correlation frequency resolved optical gating (XFROG) trace, i.e. the sum frequency spectrum vs the pump-probe delay, is reported in Fig.~\ref{Figura6}. The width of the cross correlation trace at different wavelengths ($\lambda_{XFROG}$) sets the effective temporal resolution of the transient response at a given probe wavelength ($\lambda_{probe}$), according to the energy conservation condition for the sum frequency process ($\frac{1}{\lambda_{XFROG}}=\frac{1}{\lambda_{probe}}+\frac{1}{\lambda_{pump}}$). In Fig.~\ref{Figura6} we report a single time-trace at fixed wavelength, i.e., $\lambda_{probe}$=930~nm. Since the pump spectrum is centered at $\lambda_{pump}$=590~nm, the time-trace at $\lambda_{probe}$=930~nm has to be compared to the cross-correlation trace at $\lambda_{XFROG}$=360~nm whose temporal width $\Delta\tau_{XFROG}$ is estimated to be 19$\pm$2~fs. The rise-time of the time trace is longer than the convolution (grey line) between the FROG trace (XFROG) and an Heaviside function (H), indicating a finite delay in the maximum variation of $\delta R(\omega$,t)/$R_{eq}(\omega)$. 

\section{Relaxation dynamics in the $t$-$J$ model}

\subsection{Finite-size scaling}

\begin{figure}[b]
\centering
\includegraphics[clip,width=0.32\textwidth] {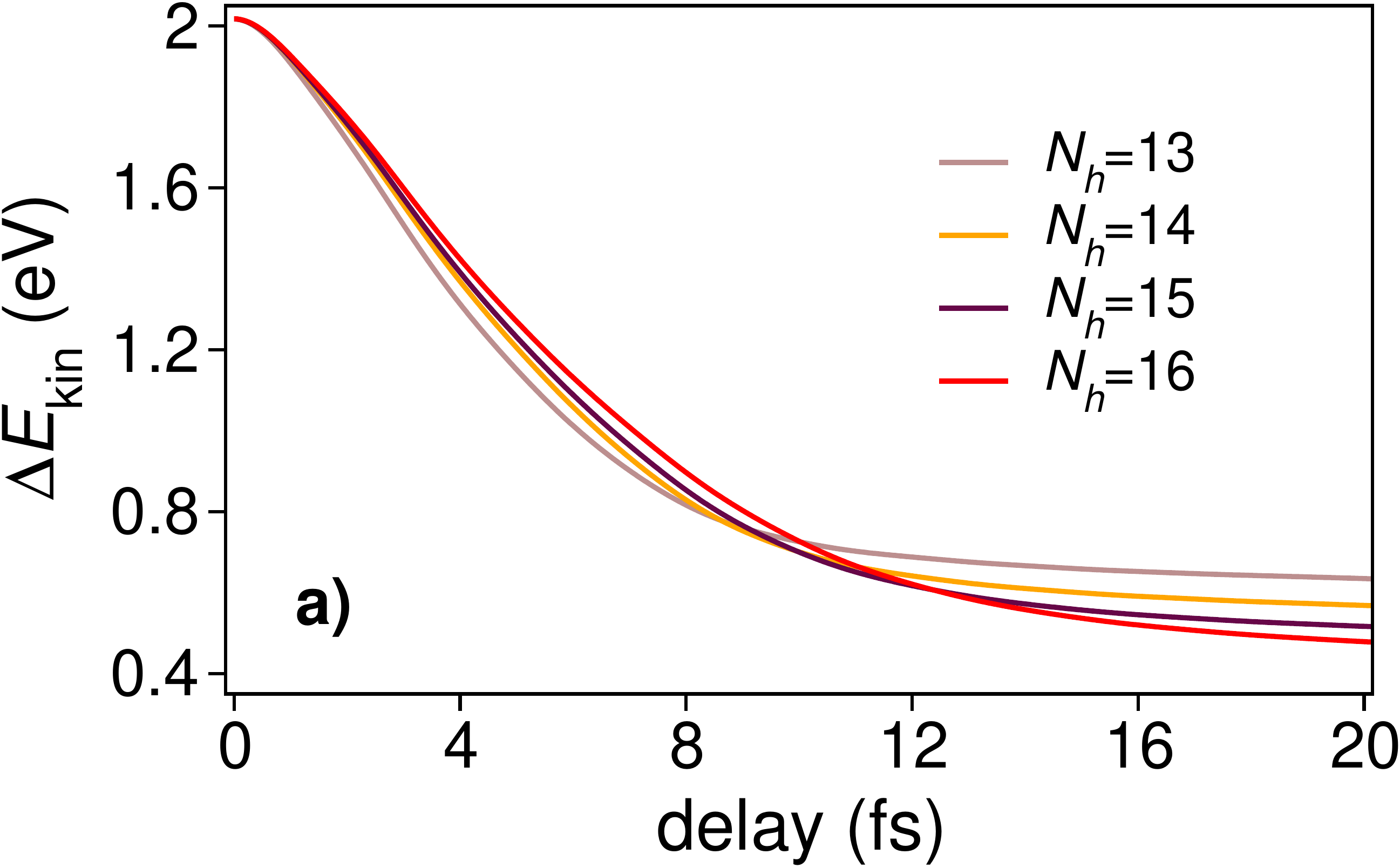}
\includegraphics[clip,width=0.32\textwidth] {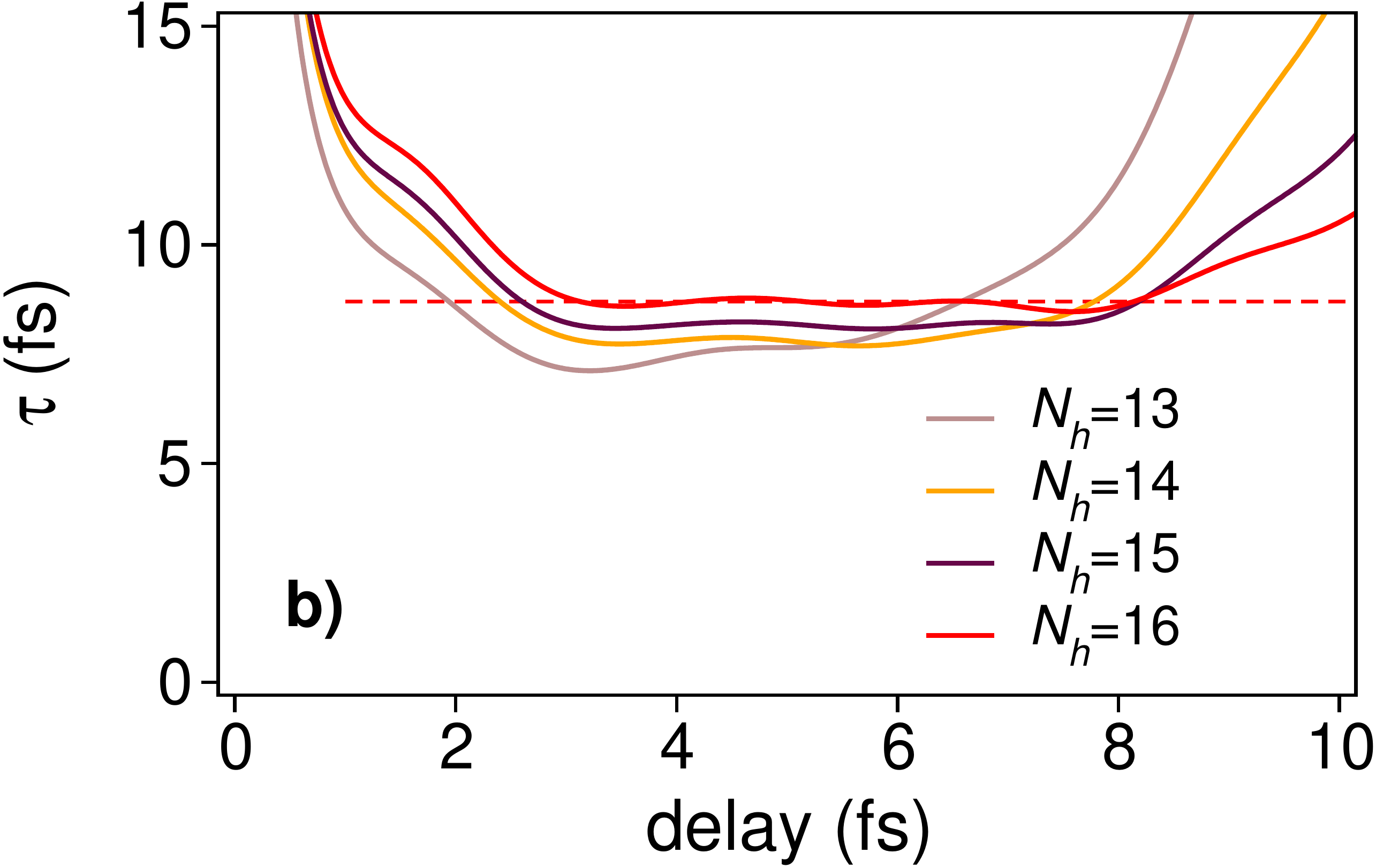}
\includegraphics[clip,width=0.32\textwidth] {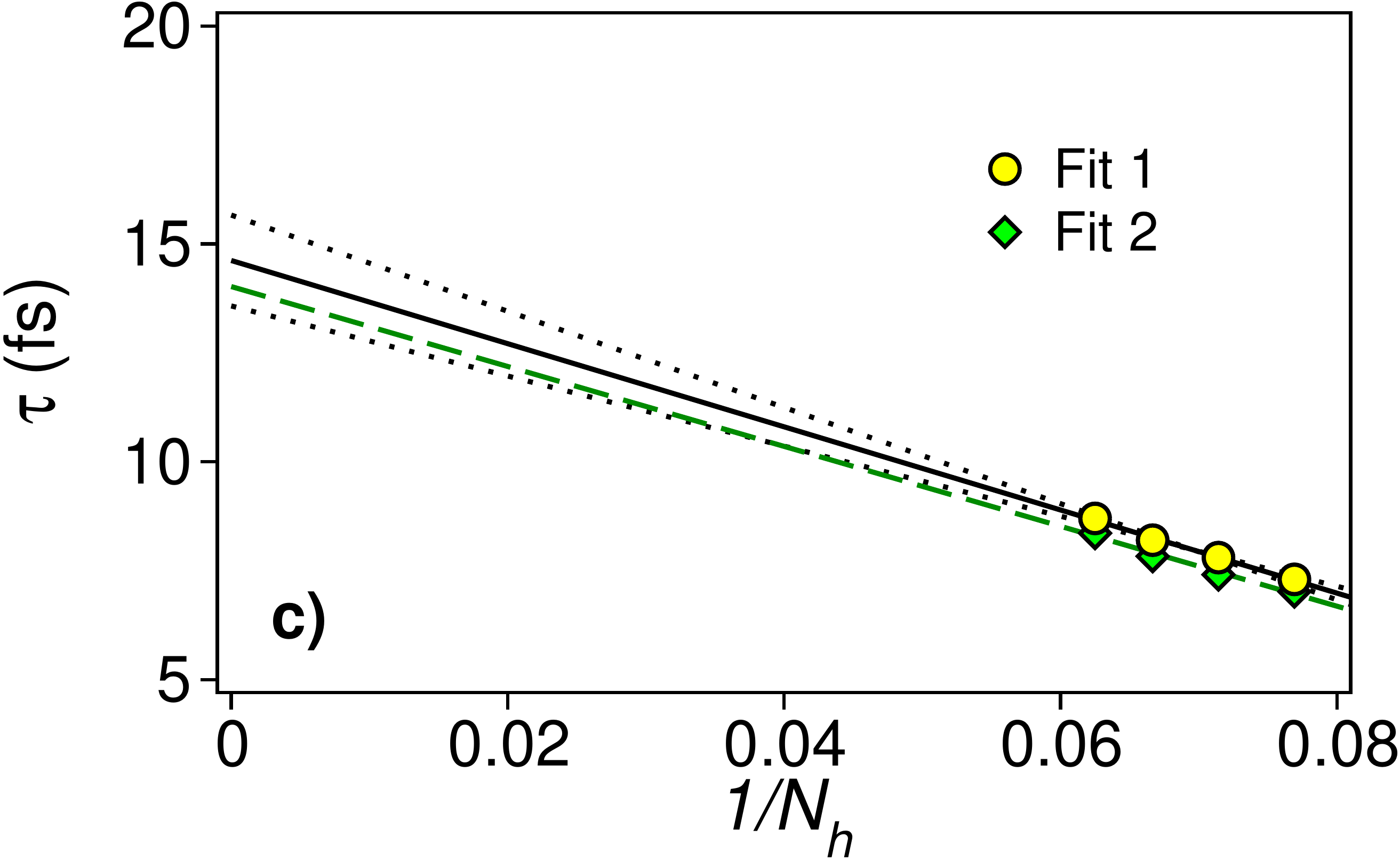}
\caption{\textbf{
a) Kinetic energy of a photo-excited hole, $\Delta E_{\rm kin}(t)$, for different sizes of the functional space, given by the parameter $N_h$.
b) The function $\tau(t) = - \Delta E_{\rm kin}(t)/\Delta \dot{E}_{\rm kin}(t)$, where $\Delta \dot{E}_{\rm kin}(t)$ denotes time-derivative, for the same system sizes.
Horizontal dashed line indicates the plateau for $N_h=16$ where $\tau(t)$ becomes time-independent.
c) $\tau$ vs $1/N_h$ for two different fitting procedures.
In Fit 1, we extract $\tau^{(1)}(N_h)$ from a time-independent plateau of $\tau(t)$, as indicated in panel b).
In Fit 2, we extract $\tau^{(2)}(N_h)$ from a fitting function $f_2(t)= A_0 \exp{(-\sqrt{at^2 + b^2}+b)}$.
Details are described in the main text.
Both fitting procedures yield a very similar relaxation time, $\tau_{tJ}\approx 15$~fs.
Dotted lines represent extrapolation for Fit 1, with fitting parameters shifted by 3 standard deviations relative to the mean values.
}}
\label{Figura_tj}
\end{figure}

We now describe the procedure adopted to extract the relaxation time from real-time numerical simulations of the $t$-$J$ model far from equilibrium.
We perform a finite-size scaling with respect to the parameter $N_h$ (see Methods) which defines the size of the functional space.
For a fixed value of $N_h$, we extract the relaxation time $\tau(N_h)$ from the time evolution of the kinetic energy $\Delta E_{\rm kin}(t)$ of a photo-excited hole.
We show  $\Delta E_{\rm kin}(t)$ for different $N_h$ in Fig.~\ref{Figura_tj}a.
Once $\tau(N_h)$ is obtained, we extrapolate $\tau_\infty = \tau(N_h\to\infty)$ with respect to the scaling with $1/N_h$.
This procedure is analogous to the finite-size scaling in the conventional exact diagonalization where the size of the Hilbert space is determined by the lattice size $L$.
In both cases, Hilbert spaces grow exponentially with either $N_h$ or $L$.
In the following, we fix the parameter values to the ones discussed in the main part of the manuscript, i.e, we set $J/t_h=0.3$, $N_{\rm box}=4$ (defining the spatial extend of local antiferromagnetic excitations; see Methods) and the initial quench energy $\Delta E_{\rm kin}=2$~eV. 

We obtain the relaxation time from two independent and partially complementary procedures.
We start with a simple exponential fit $f_1(t) = A \exp{(-t/\tau^{(1)})}$ of the kinetic energy of the photo-excited hole.
This is consistent with the fitting procedure of the experimental data for $\delta \gamma_\infty(t)$, shown in Fig.~1 of the main part of the manuscript.
However, some care is required when fitting the numerical data with this ansatz.
The ansatz is clearly not correct in the very first stage of relaxation, i.e., in the time of the order of a single hopping process, $t\sim 1$~fs.
In addition, in the asymptotic limit where the majority of the kinetic energy is already released to local antiferromagnetic excitations, a finite functional space limits the spreading of these excitation through the plaquette, as seen for different values of $\Delta E_{\rm kin}(t)$ around $t\sim 20$~fs in Fig.~\ref{Figura_tj}a.
We therefore calculate in Fig.~\ref{Figura_tj}b the function $\tau(t) = - \Delta E_{\rm kin}(t)/\Delta \dot{E}_{\rm kin}(t)$, which gives an instantaneous relaxation time.
In the case of the exponential function $f_1(t)$, it simply gives $\tau(t)=\tau^{(1)}$.
Remarkably, we observe a plateau of $\tau(t)$ in the time interval $2$~fs~$< t <10$~fs in Fig.~\ref{Figura_tj}b, which indicates that a large part of the relaxation process can be to a very good approximation described by a simple exponential function.
The time-independent regime of $\tau(t)$ in Fig.~\ref{Figura_tj}b can only be detected for large functional spaces (obtained for $N_h=13,14,15,16$), which contain over $10^6$ states.
Most importantly, the regime of exponential relaxation with $\tau(t)={\it const}$ extends to longer times for larger values of $N_h$.
The values of $\tau^{(1)}(N_h)$ obtained from the plateau of $\tau(t)$ are shown in Fig.~\ref{Figura_tj}c (circles).
In the latter Figure, the linear extrapolation with respect to $1/N_h$ provides a very accurate fit for the available data and yields $\tau_\infty^{(1)}=14.6$~fs.

We complement the fitting procedure with a fitting function $f_2(t) = A_0 \exp{(-\sqrt{at^2 + b^2}+b)}$, which, unlike $f_1(t)$, also captures the time evolution of the kinetic energy at the very short time, of the order of $t\sim 1$~fs.
Moreover, it obeys the time-reversal symmetry of the investigated quantum-mechanical system S\cite{hamerla2013}.
It gives $\dot{f}_2(t)=0$ at $t=0$, consistent with our numerical data, as well as it describes a simple exponential decay $f_2(t) \sim \exp{(-t/\tau^{(2)})}$ at long times, where $\tau^{(2)}=1/\sqrt{a}$.
We cut the numerical data at the time $t^*$ when $\tau(t)$ starts to deviate from the plateau shown in Fig.~\ref{Figura_tj}b, which depends on $N_h$ and occurs typically between $t^*\in$ [6~fs,~8~fs].
The rest of the data is then fitted by $f_2(t)$.
The resulting values of $\tau^{(2)}=1/\sqrt{a}$ are shown as diamonds in Fig.~\ref{Figura_tj}c.
We again extrapolate $\tau^{(2)}$ with respect to $1/N_h$ for large functional spaces which contain over $10^6$ states (solid line in Fig.~\ref{Figura_tj}c) and we get $\tau_\infty^{(2)}=14.1$~fs.

In both fitting procedures described above, the values of $\tau_\infty^{(1)}$ and $\tau_\infty^{(2)}$ are quantitatively very similar.
We therefore take $\tau_{tJ} = 15$~fs throughout the manuscript to quantify the relaxation time in the $t$-$J$ model.
In addition, even the bare relaxation time obtained for a large but finite Hilbert space ($\tau=9$~fs for $N_h=16$) yields the characteristic relaxation time scale of the order of 10~femtoseconds.
This extremely fast relaxation due to the charge-spin interaction is the main message of our numerical simulations.

\subsection{The role of a direct carrier-carrier interaction in the lower Hubbard band}
\begin{figure}[b]
\centering
\includegraphics[clip,width=0.32\textwidth]{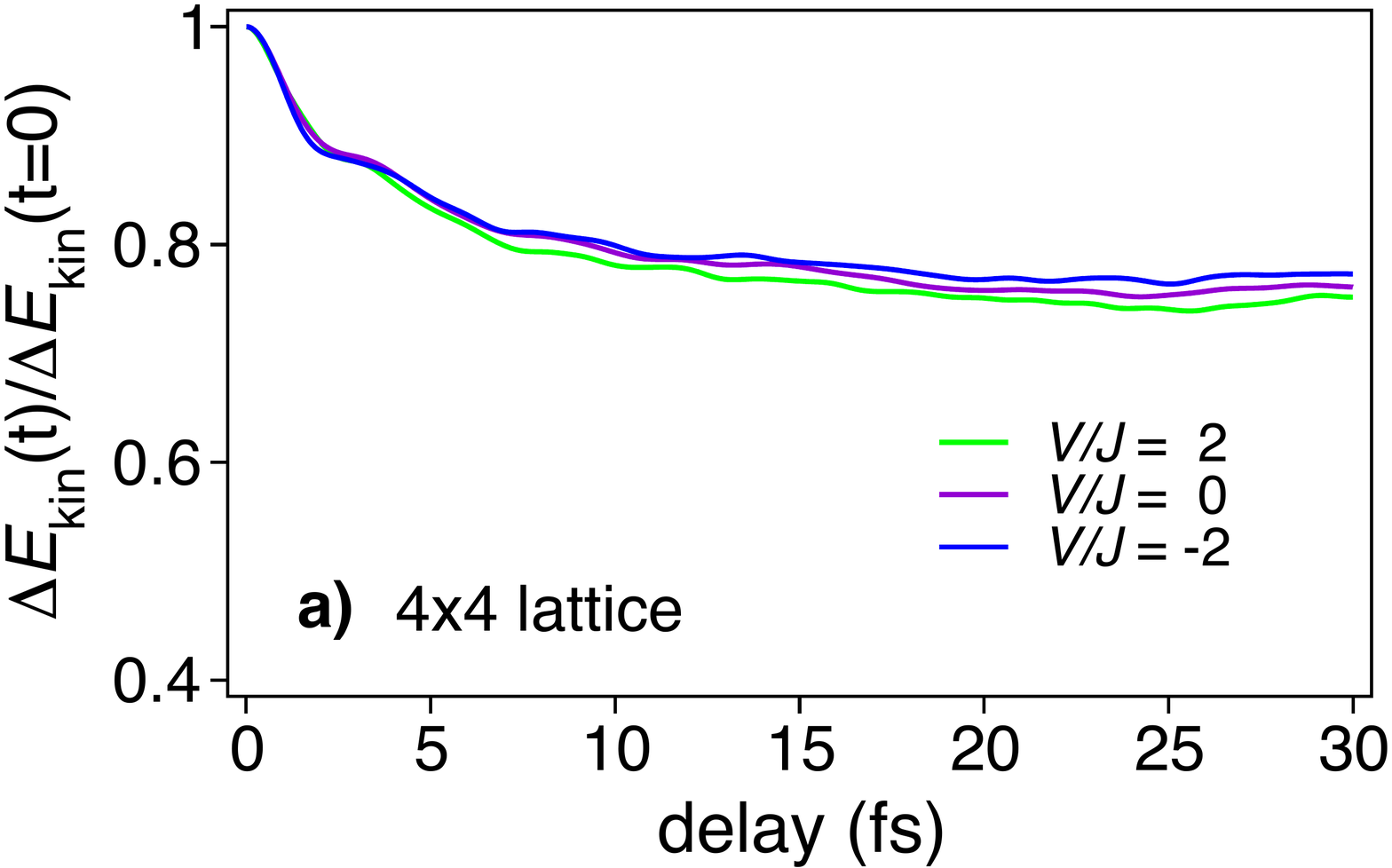}
\hspace*{1cm}
\includegraphics[clip,width=0.32\textwidth]{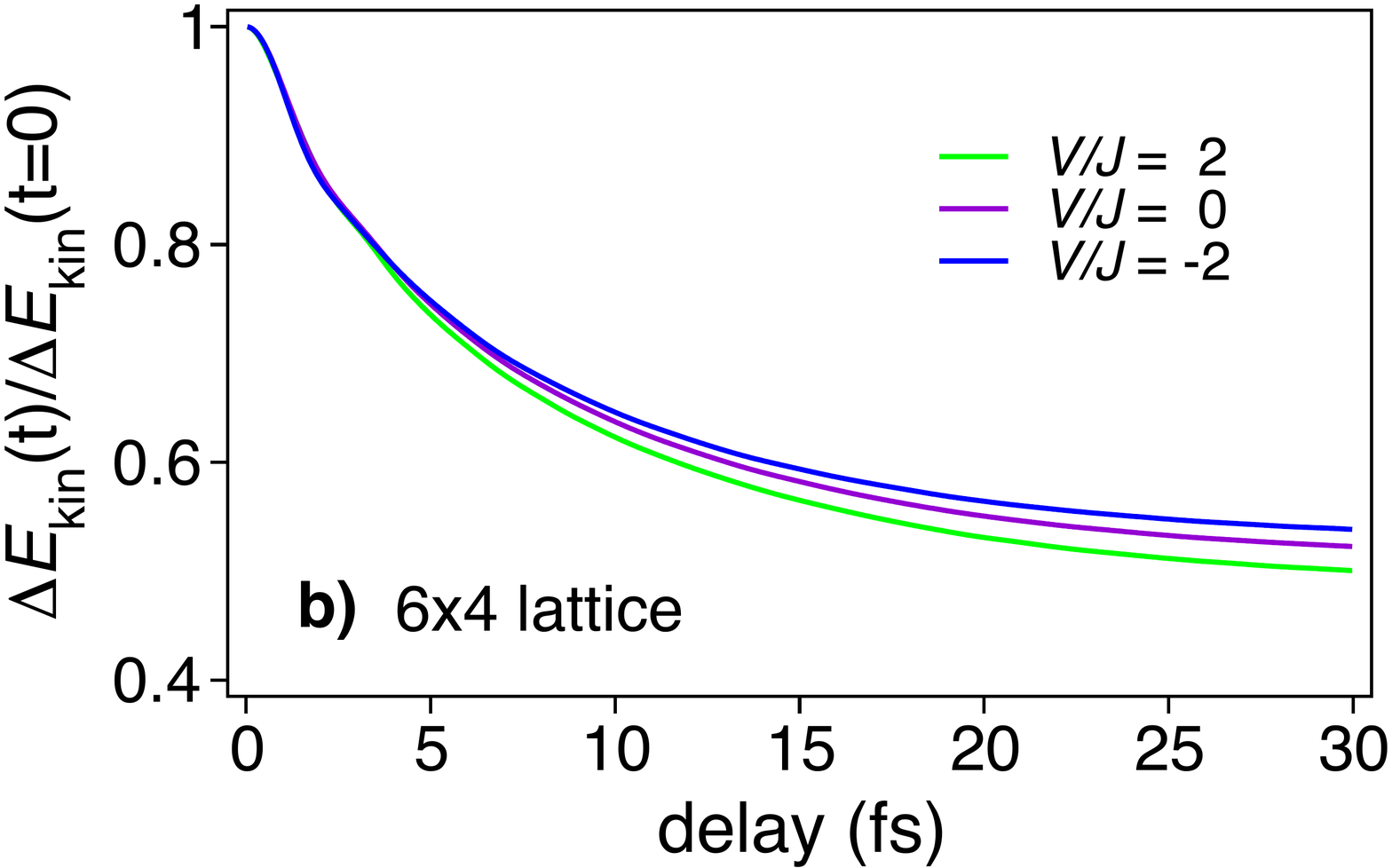}
\caption{\textbf{
Kinetic energy $\Delta E_{\rm kin}(t)/\Delta E_{\rm kin}(t$=$0)$ in the system of two carriers, for different values of the direct carrier-carrier interaction $V$, see Eq.~(\ref{def_tjv}).
The other parameters are the same as in Fig. 2 of the main text, i.e., $J/t_h=0.3$, $t_h=360$ meV and $\Delta E_{\rm kin}(t=0)=2$ eV. Results have been obtained by exact diagonalization on a) 4$\times$4 and b) 6$\times$4 square lattice using periodic boundary conditions.
}}
\label{Figura_ed}
\end{figure} 


Motivated by the extremely low density of the photo-excited carriers (shortly carriers) we have so far studied a single carrier in the $t$-$J$ model,
thus neglecting carrier-carrier interactions (see the discussion in Methods).
In this subsection we provide an additional support for this picture and explicitly show that this  interaction does not influence the ultrafast relaxation of highly excited carriers at least for parameters which are relevant for the present experiment.
The direct interaction between carriers in the lower Hubbard band is already captured in the original $t$-$J$ model, however it is not obvious how to single out effects originating solely from this interaction. 
The first-guess solution to this problem would be to study how the relaxation dynamics depends on the concentration of carriers. 
Unfortunately such an approach, apart from being technically very demanding, would also be conceptually unjustified.
Changing the carrier concentration namely modifies the magnetic background that consequently  influences the coupling between carriers and the magnetic excitations.
In order to overcome this problem we have extended the original $t$-$J$  Hamiltonian by an additional term describing a direct carrier-carrier interaction with a potential $V$,
\begin{equation} \label{def_tjv}
H_{tJ} \rightarrow H_{tJV}=H_{tJ}+V\sum_{\langle i,j\rangle} \tilde{n}_i \tilde{n}_j,
\end{equation}
where the $t$-$J$ Hamiltonian $H_{tJ}$ as well as the electron-number operator $\tilde{n}_j $ have been explained in Eq.~(1) of Methods.
After such an extension the interaction between a carrier and magnetic excitations still depends only on $J$, while the carrier-carrier interaction can be tuned by changing either
$J$ or $V$ S\cite{bonca2012}.
Below we demonstrate that for a fixed $J$ the relaxation dynamics barely depends on $V$, indicating that the strength of the carrier-carrier interaction has no significant influence on the relaxation time.
It is important to stress that we do not assign any physical relevance to the additional interaction term in Eq.~(\ref{def_tjv}).
It should only be considered as an auxiliary interaction introduced to establish the experimental relevance of results obtained for a single charge carrier in the $t$--$J$ model.

Figure~\ref{Figura_ed} shows the relaxation of the kinetic energy for two charge carriers propagating on 4$\times$4 and 6$\times$4 plaquettes solved by exact diagonalization (ED).
Note that the carrier concentration in the former case is close to the hole concentration in the optimally doped cuprates. 
We use ED since the two-carrier system on larger plaquettes solved by diagonalization in the limited functional space converges much slower than the corresponding single-carrier system. 
As discussed in the Methods, clusters with $N<30$ sites solved by ED represent too small systems to account for a complete relaxation within the experimental conditions, and therefore the relaxation times should not be compared quantitatively to the data shown in Fig.~\ref{Figura_tj}.
Nevertheless, the main message from the results in Fig.~\ref{Figura_ed} does not concern the quantitative results for the relaxation time but it rather tests the sensitivity of the relaxation time on the coupling between carriers as modeled by $V$.
We have computed  the relaxation dynamics in the presence of  attractive as well as repulsive potential $V$  with a magnitude that is sufficiently large to overcome the carrier-carrier interaction present in the original $t$--$J$ model.
Results are quite independent of $V$, and different panels in Fig.~\ref{Figura_ed} demonstrate that the same behavior persists for various sizes of the clusters.
We thus conclude that the direct carrier-carrier interaction in the lower Hubbard band has a inconsequential influence on  the ultrafast primary relaxation in the time-window $t\lesssim 15~{\rm fs}$. However, we expect that this interaction may become important for larger doping (and in particular for large fluence), when magnetic correlations diminish while the density of carriers becomes  larger.
Finally we stress that the direct carrier-carrier interaction between particles in the lower Hubbard band
should not be confounded with the original Hubbard repulsion $U$ in the full (unprojected) Hilbert space, since the latter interaction is actually the physical origin behind the coupling between the charge carriers and  the antiferromagnetic fluctuations.

\section{Dynamical mean-field theory (DMFT)}
\begin{figure}[t]
\includegraphics[keepaspectratio,clip,width=0.70\textwidth]{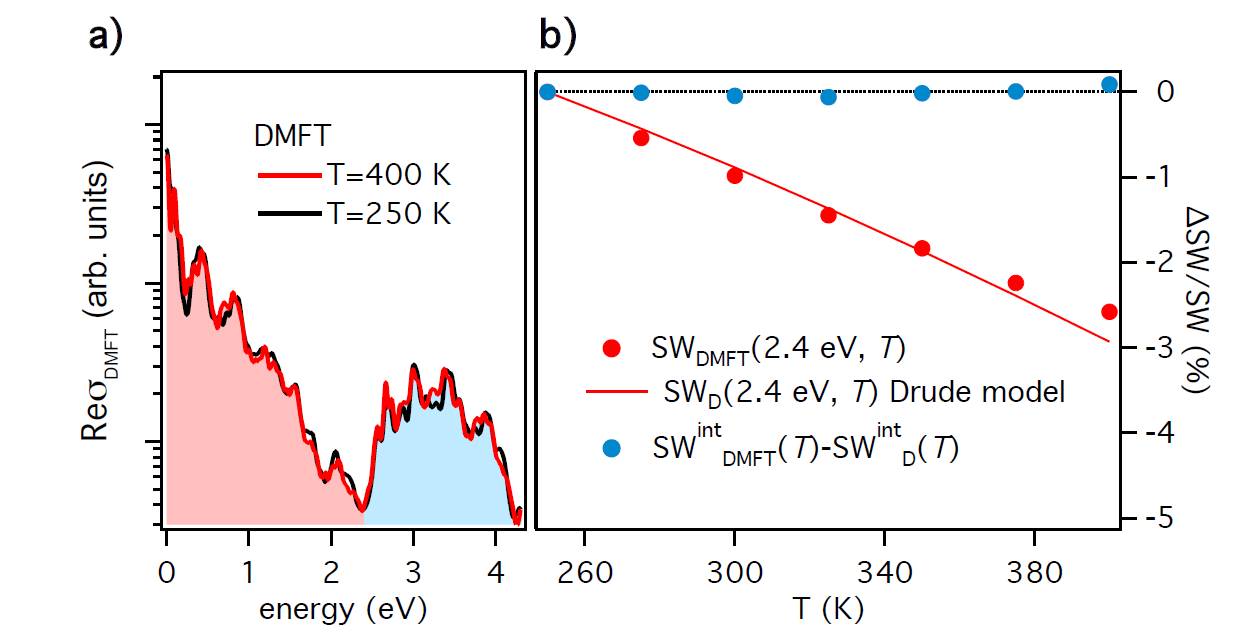}
\caption{\textbf{a) The optical conductivity of the single band Hubbard Model calculated by DMFT is reported for T=250 and 400 K and $U$=3 eV. A broad Drude peak extends up to 2.4 eV, whereas the optical properties are dominated by interband transitions at the $U$ energy scale. b) The relative variation of the optical spectral weight, i.e., $\Delta$SW/SW=SW$_i$($\Omega_c$,$T$)-SW$_i$($\Omega_c$,$250$K)/SW$(\Omega_c$,$250$K), as a function of the temperature $T$. The temperature-dependent intraband spectral weight SW$_i$(2.4 eV,$T$)=$\int_0^{2.4 {\rm eV}}$Re$\sigma_i(\omega,T) d\omega$ is calculated for the DMFT optical conductivity ($i$=$\rm DMFT$, red dots) and for an extended Drude model ($i$=$\rm D$). The interband spectral weight is calculated as the difference between SW$^{\rm int}_{\rm DMFT}$($T$)=$\int_{2.4 {\rm eV}}^{4.2 {\rm eV}}$Re$\sigma_{\rm DMFT}(\omega,T) d\omega$ and the intrinsic spectral weight of the underlying Drude tail, i.e., SW$^{\rm int}_{\rm D}$($T$)=$\int_{2.4 {\rm eV}}^{4.2 {\rm eV}}$Re$\sigma_{\rm D}(\omega,T) d\omega$.}}
\label{figDMFT}
\end{figure}
The role of short-range AF correlations as the effective bosons mediating the electron dynamics in doped cuprates is also supported by the Dynamical Mean Field Theory (DMFT) solution of the Hubbard model S\cite{dmft}.
Even if an accurate characterization of the phase diagram of the copper-oxide planes of the cuprates requires cluster extensions of DMFT S\cite{cluster}, the single-site DMFT accurately reproduces the evolution of the optical spectral weight as a function of both doping S\cite{comanac,toschi} and temperature S\cite{nicoletti}.
We consider a two-dimensional Hubbard model with $t$=0.25 eV, $t'$=-0.05 eV and U=3eV, and we solve the local quantum dynamics using finite-temperature exact diagonalization S\cite{cdg} approximating the bath with 8 levels. The frequency-dependent optical conductivity is directly obtained from the knowledge of the single particle Green's functions and of the current vertex.
Using DMFT, we calculate the temperature-dependent optical conductivity, $\sigma(\omega,T)$, of a hole-doped material whose properties are assumed to be described by a generic single-band Hubbard model. DMFT  is used to compute $\sigma(\omega,T)$, shown in Fig.~\ref{figDMFT}a, fully retaining both the physics of the high-energy transitions at the energy $U$ and that of the short-range antiferromagnetic correlations at the energy scale $J$. 
At an intermediate hole concentration ($p$=0.16), $\sigma(\omega,T)$ is characterized by a broad Drude-like part ($\hbar\omega$$<$2.4 eV) and distinct interband transitions in the 2.4-4.2 eV energy range, that account for the localized excitations from the lower to the upper Hubbard band. As shown in Fig.~\ref{figDMFT}a, the main effect of a temperature increase from 250 K (black line) to 400 K (red line) is the overall broadening of the Drude peak with almost no change of spectral weight at the energy $U$. Quantitatively, this is shown by calculating the finite-cutoff spectral weight, i.e.: 
\begin{equation}
SW(\Omega_c,T)=\int_0^{\Omega_c}\mathrm{Re}\sigma(\omega,T) d\omega
\end{equation}
as a function of the temperature $T$. In the simplest version of the Drude model, in which the scattering rate is frequency-independent, the approximation: 
\begin{equation}
SW(\Omega_c,T)\simeq \frac{\omega^2_{p}}{8}\left[1-2\frac{\gamma(T)}{\pi \Omega_c}\right]
\label{SW}
\end{equation}
is valid in the limit $\Omega_c$$\gg$$\gamma$. Therefore, the temperature dependance of SW($\Omega_c$,$T$) directly reflects that of $\gamma(T)$. 
In the Fig.~\ref{figDMFT}b, we show that SW($\Omega_{1}$=2.4eV,$T$), calculated from the output of DMFT, is in perfect agreement with that expected for a simple Drude model (SW$_D$($\Omega$,$T$)), provided a suitable effective electron-boson scattering rate is introduced. The quantity SW$_D$(2.4 eV,$T$), reported in Fig.~\ref{figDMFT}b as a red line, has been obtained by replacing in Eq.~\ref{SW} the asymptotic temperature-dependent scattering rate, $\gamma_{\infty}(T)$ calculated through Eqs.~\ref{selfenergy}, \ref{kernel} and \ref{scatteringrate}. The bosonic function $I^2 \chi(\Omega)$ is taken from Ref. \onlinecite{DalConte2012}. The calculated SW$_D$(2.4 eV,$T$) is robust against modifications of the details of $I^2 \chi(\Omega)$ in the 0-500 meV energy range.
Furthermore, the slope of SW($\Omega_2$=4.2eV,$T$) exactly scales with the ratio $\Omega_1$/$\Omega_2$ (see Eq.~\ref{SW}) finally demonstrating that, at sufficiently high temperatures and hole concentrations, the dynamics of the charge carriers can be entirely described through an effective electron-boson coupling on the scale of $J$ (no other energy scale is present in the model), without any change of spectral weight of high-energy Mott-like excitations at $U$.



\begin{thebibliography}{1}

\bibitem{Lee2006}
Lee, P.A., Nagaosa, N. and Wen, X.-G. Doping a Mott insulator: Physics of high-temperature superconductivity. \textit{Rev. Mod. Phys.} \textbf{78}, 17 (2006).

\bibitem{Anderson2007}
Anderson, P. W. Is there glue in cuprate superconductors? \textit{Science} \textbf{316}, 1705 (2007).

\bibitem{Scalapino2012}
Scalapino, D.J. A common thread: The pairing interaction for unconventional superconductors. \textit{Rev. Mod. Phys.} \textbf{84}, 1383 (2012).

\bibitem{Monthoux2007}
Monthoux, P., Pines, D., and Lonzarich, G. G. Superconductivity without phonons. \textit{Nature} \textbf{450}, 1177 (2007).

\bibitem{Orenstein2012}
Orenstein, J., Ultrafast spectroscopy in quantum materials. \textit{Physics Today} \textbf{65}, 44-50 (2012).

\bibitem{Brida2010}
Brida, D., \textit{et al.} Few-optical-cycle pulses tunable from the visible to the mid-infrared by optical parametric amplifiers. \textit{Journal of Optics} \textbf{12}, 013001 (2010).

\bibitem{dalconte12}
Dal Conte, S., \textit{et al.} Disentangling the electronic and phononic glue in a high-T$_{c}$ superconductor. \textit{Science} \textbf{335}, 1600-1603 (2012).

\bibitem{Eisaki2004}
Eisaki, H., \textit{et al.} Effect of chemical inhomogeneity in bismuth-based copper oxide superconductors. \textit{Phys. Rev. B} \textbf{69}, 064512 (2012).

\bibitem{Cilento2014}
Cilento, F., \textit{et al.} Photo-enhanced antinodal conductivity in the pseudogap state of high-T$_{c}$ cuprates. \textit{Nat. Commun.} \textbf{5}, 4353 (2014).

\bibitem{Heumen2009}
Van Heumen, E., \textit{et al.} Optical determination of the relation between the electron-boson coupling function and the critical temperature in high-T$_{c}$ cuprates. \textit{Phys. Rev. B} \textbf{79}, 184512 (2009).

\bibitem{Novelli2013}
Novelli, F., \textit{et al.} Witnessing the formation and relaxation of dressed quasi-particles in a strongly correlated electron system. \textit{Nat. Commun.} \textbf{5}, 5112 (2014).

\bibitem{perfetti07}
Perfetti, L., \textit{et al.} Ultrafast Electron Relaxation in Superconducting Bi$_{2}$Sr$_{2}$CaCu$_{2}$O$_{8+\delta}$. \textit{Phys. Rev. Lett.} \textbf{99}, 197001 (2007).

\bibitem{gadermaier10}
Gadermaier, C., \textit{et al.} Electron-phonon coupling in high-temperature cuprate superconductors determined from electron relaxation rates. \textit{Phys. Rev. Lett.} \textbf{105}, 257001 (2010).

\bibitem{Dahm2009}
Dahm, T., \textit{et al.} Strength of the spin-fluctuation-mediated pairing interaction in a high-temperature superconductor. \textit{Nature Phys.} \textbf{5}, 217-221 (2009).

\bibitem{letacon2011}
Le Tacon, M., \textit{et al.} Intense paramagnon excitations in a large family of high-temperature superconductors. \textit{Nature Phys.} \textbf{7}, 725-730 (2011).

\bibitem{Fujita2012}
Fujita, M., \textit{et al.} Progress in Neutron Scattering Studies of Spin Excitations in High-Tc Cuprates. \textit{J. Phys. Soc. Jpn.} \textbf{81}, 011007 (2012).

\bibitem{dean2013}
Dean, M. P. M., \textit{et al.} High-Energy Magnetic Excitations in the Cuprate Superconductor Bi$_{2}$Sr$_{2}$CaCu$_{2}$O$_{8}$: Towards a Unified Description of Its Electronic and Magnetic Degrees of Freedom. \textit{Phys. Rev. Lett.} \textbf{110}, 147001 (2013).

\bibitem{letacon2013}
Le Tacon, M., \textit{et al.} Dispersive spin excitations in highly overdoped cuprates revealed by resonant inelastic x-ray scattering. \textit{Phys. Rev. B} \textbf{88}, 020501 (2013).

\bibitem{dean2013b}
Dean, M. P. M., \textit{et al.} Persistence of magnetic excitations in La$_{2-x}$Sr$_{x}$CuO$_{4}$ from the undoped insulator to the heavily overdoped non-superconducting metal. \textit{Nature Mater.} \textbf{12}, 1019-1023 (2013).

\bibitem{Singla2013}
Singla, R., \textit{et al.} Photoinduced melting of the orbital order in La$_{0.5}$Sr$_{1.5}$MnO$_{4}$ measured with 4- fs laser pulses. \textit{Phys. Rev. B} \textbf{88}, 075107 (2013).

\bibitem{mierzejewski2011}
Mierzejewski, M., Vidmar, L., Bon\v{c}a, J. and Prelov\v{s}ek, P. Nonequilibrium quantum dynamics of a charge carrier doped into a mott insulator. \textit{Phys. Rev. Lett.} \textbf{106}, 196401 (2011).

\bibitem{Golez2013}
Gole\v{z}, D., Bon\v{c}a, J., Mierzejewski, M. and Vidmar, L. Mechanism of Ultrafast Relaxation of a Photo-Carrier in Antiferromagnetic Spin Background. \textit{Phys. Rev. B} \textbf{89}, 165118 (2014).

\bibitem{Kong2001}
Kong, Y., Dolgov, O., Jepsen, O. and Andersen, O. Electron-phonon interaction in the normal and superconducting states of MgB$_{2}$. \textit{Phys. Rev. B} \textbf{64}, 020501 (2001).

\bibitem{DellaValle2012}
Della Valle, G., Conforti, M., Longhi, S., Cerullo, G. and Brida, D. Real-time optical mapping of the dynamics of nonthermal electrons in thin gold films. \textit{Phys. Rev. B} \textbf{86}, 155139 (2012).

\bibitem{ElsayedAli1987}
Elsayed-Ali, H., Norris, T., Pessot, M. and Mourou, G. Real-time optical mapping of the dynamics of nonthermal electrons in thin gold films. \textit{Phys. Rev. B} \textbf{58}, 1212-1215 (1987).

\bibitem{Allen1975}
Allen, P. and Dynes, R. Transition temperature of strong-coupled superconductors reanalyzed. \textit{Phys. Rev. B} \textbf{12}, 905-922 (1975).

\bibitem{Allen1987}
Allen, P. Theory of thermal relaxation of electrons in metals. \textit{Phys. Rev. Lett.} \textbf{59}, 1460-1463 (1987).

\bibitem{Eckstein2014}
Eckstein, M. and Werner, P. Ultra-fast photo-carrier relaxation in Mott insulators with short- range spin correlations. arXiv:1410.3956 (2014).

\bibitem{Hinks2009}
Hinks, D. G. Iron arsenide superconductors: What is the glue? \textit{Nature Phys.} \textbf{5}, 386-387 (2009).

\bibitem{Mazin2010}
Mazin, I. I. Superconductivity gets an iron boost. \textit{Nature} \textbf{464}, 183-186 (2010).

\bibitem{Brida2009}
Brida, D. \textit{et al.} Generation of 8.5 fs pulses at 1.3 $\mu$m for ultrabroadband pump-probe spectroscopy. \textit{Optics Express} \textbf{464}, 12510 (2009).

\bibitem{Ono2003}
Ono, S. and Ando, Y. Evolution of the resistivity anisotropy in Bi$_{2}$Sr$_{2-x}$La$_{x}$CuO$_{6+\delta}$ single crystals for a wide range of hole doping. \textit{Phys. Rev. B} \textbf{67}, 104512 (2003).

\bibitem{Peets2007}
Peets, D. \textit{et al.} Tl$_{2}$Ba$_{2}$CuO$_{6+\delta}$ brings spectroscopic probes deep into the overdoped regime of the high-T$_c$ cuprates. \textit{New J. Phys.} \textbf{9}, 28 (2007).

\bibitem{Zhigadlo2010}
Zhigadlo, N. D. \textit{et al.} Influence of Mg deficiency on crystal structure and superconducting
properties in MgB$_2$ single crystals. \textit{Phys. Rev. B} \textbf{81}, 054520 (2010).

\bibitem{Bonca2007}
Bon\v{c}a, J., Maekawa, S. and Tohyama, T. Numerical approach to the low-doping regime of the t-J model. \textit{Phys. Rev. B} \textbf{76}, 035121 (2007).

\bibitem{park1986}
Park, T. J. and Light, J. C. Unitary quantum time evolution by iterative Lanczos reduction. \textit{J. Chem. Phys.} \textbf{85}, 5870-5876 (1986).


\end{thebibliography}

\begin{thebibliography}{1}

\renewcommand*{\citenumfont}[1]{S#1}
\renewcommand*{\bibnumfmt}[1]{[S#1]}



\bibitem{Giannetti2011}
Giannetti, C. \textit{et. al.} Revealing the high-energy electronic excitations underlying the onset of high-temperature superconductivity in cuprates. \textit{Nat. Commun.} \textbf{3}, 253 (2011).

\bibitem{Coslovich2013}
Coslovich, G. \textit{et. al.} Competition between the pseudogap and superconducting states of Bi${}_{2}$Sr${}_{2}$CaCu${}_{2}$O${}_{8+\delta}$ single crystals revealed by ultrafast broadband optical reflectivity. \textit{Phys. Rev. Lett} \textbf{110}, 107003 (2013).

\bibitem{DalConte2012}
Dal Conte, S. \textit{et. al.} Disentangling the Electronic and Phononic Glue in a High-$T_c$ Superconductor. \textit{Science} \textbf{335}, 1600 (2012).

\bibitem{Allen1971}
Allen, P.B., Electron-phonon effects in the infrared properties of metals. \textit{Phys. Rev. B} \textbf{3}, 305 (1971).

\bibitem{VanHeumen2009}
Van Heumen, E. \textit{et. al.} Optical determination of the relation between the electron-boson coupling function and the critical temperature in high-T$_{c}$ cuprates \textit{Phys. Rev. B} \textbf{79}, 184512 (2009).

\bibitem{Carbotte2011}
Carbotte, J.P \textit{et. al.} Bosons in high-temperature superconductors: an experimental survey. \textit{Rep. Prog. Phys.} \textbf{74}, 066501  (2011).

\bibitem{Guritanu2006}
Guritanu, V. \textit{et. al.} Anisotropic optical conductivity and two colors of MgB$_{2}$. \textit{Phys. Rev. B} \textbf{73}, 104509  (2006).

\bibitem{Sun1994}
Sun, C.-K. \textit{et. al.} Femtosecond-tunable measurements of electron thermalization in gold. \textit{Phys. Rev. B} \textbf{50}, 15337   (1994).

\bibitem{DellaValle2012}
Della Valle, G. \textit{et. al.} Real-time optical mapping of the dynamics of nonthermal electrons in thin gold films. \textit{Phys. Rev. B} \textbf{86}, 155139   (2012).

\bibitem{Greger2013}
Greger, M. \textit{et. al.} Isosbestic points: How a narrow crossing region of curves determines their leading parameter dependence. \textit{Phys. Rev. B} \textbf{87}, 195140   (2013).

\bibitem{Kong2001}
Kong, Y. \textit{et. al.} Electron-phonon interaction in the normal and superconducting states of MgB$_{2}$. \textit{Phys. Rev. B} \textbf{64}, 020501(R) (2001).

\bibitem{hamerla2013}
Hamerla, S. A. and Uhrig, G. S. Interaction quenches in the two-dimensional fermionic Hubbard model.
Phys. Rev. B {\bf 89}, 104301 (2014).

\bibitem{bonca2012}
Bon\v{c}a, J., Mierzejewski, M., \& Vidmar, L. Nonequilibrium Propagation and Decay of a Bound Pair in Driven $t$-$J$ Models.
Phys. Rev. Lett. {\bf 109}, 156404 (2012)


\bibitem{dmft} 
Georges, A. \textit{et. al.}, Dynamical mean-field theory of strongly correlated fermion systems and the limit of infinite dimensions. \textit{Rev. Mod. Phys.} \textbf{68}, 13 (1996).

\bibitem{cluster} 
Maier, Th. \textit{et. al.}, Quantum cluster theories. \textit{Rev. Mod. Phys.} \textbf{77}, 1027 (2005); G. Kotliar \textit{et. al.}, Electronic structure calculations with dynamical mean-field theory. \textit{Rev. Mod. Phys.} \textbf{78}, 865 (2006).

\bibitem{comanac} Comanac, A. \textit{et. al.}, Optical conductivity and the correlation strength of high-temperature copper-oxide superconductors. \textit{Nat. Phys.} \textbf{4}, 287 (2008).

\bibitem{toschi} 
Toschi, A. and Capone, M. Optical sum rule anomalies in the cuprates: Interplay between strong correlation and electronic band structure. \textit{Phys. Rev. B} \textbf{77}, 014518 (2008).

\bibitem{nicoletti}
Nicoletti, D. \textit{et. al.}, High-Temperature Optical Spectral Weight and Fermi-liquid Renormalization in Bi-Based Cuprate Superconductors. \textit{Phys. Rev. Lett.} \textbf{105}, 077002 (2010).

\bibitem{cdg}
Capone, M. \textit{et. al.}, Solving the dynamical mean-field theory at very low temperatures using the Lanczos exact diagonalization. \textit{Phys. Rev. B} \textbf{76}, 245116 (2007).
\end{thebibliography}
\end{document}